\documentclass[floatfix,notitlepage,aps,prd,twocolumn,superscriptaddress,altaffilletter]{revtex4-1}

\usepackage{graphicx}
\usepackage{hyperref}
\usepackage{multirow}
\usepackage{subfigure}
\usepackage{siunitx}
\usepackage{color}
\usepackage{amsmath}
\newcommand{\DTO}{D$_2$O~}

\usepackage[capitalize]{cleveref}
\crefname{figure}{fig.}{figs.}
\crefname{table}{table}{tables.}
\crefname{equation}{eqn.}{eqns.}
\crefname{section}{sec.}{secs.}

\begin{document}

\title{Measurement of neutron production in atmospheric neutrino interactions at the Sudbury Neutrino Observatory}

%
\newcommand{\dec}{Deceased}
\newcommand{\alta}{Department of Physics, University of 
Alberta, Edmonton, Alberta, T6G 2R3, Canada}
\newcommand{\chicago}{Department of Physics, University of 
Chicago, Chicago IL} 
\newcommand{\ubc}{Department of Physics and Astronomy, University of 
British Columbia, Vancouver, BC V6T 1Z1, Canada}
\newcommand{\bnl}{Chemistry Department, Brookhaven National 
Laboratory,  Upton, NY 11973-5000}
\newcommand{\carleton}{Ottawa-Carleton Institute for Physics, Department of Physics, Carleton University, Ottawa, Ontario K1S 5B6, Canada}
\newcommand{\carletona}{Department of Physics, Carleton University, Ottawa, Ontario, Canada}
\newcommand{\uog}{Physics Department, University of Guelph,  
Guelph, Ontario N1G 2W1, Canada}
\newcommand{\lu}{Department of Physics and Astronomy, Laurentian 
University, Sudbury, Ontario P3E 2C6, Canada}
\newcommand{\lbnl}{Institute for Nuclear and Particle Astrophysics and 
Nuclear Science Division, Lawrence Berkeley National Laboratory, Berkeley, CA 94720-8153}
\newcommand{\lbla}{ Lawrence Berkeley National Laboratory, Berkeley, CA}
\newcommand{\lanl}{Los Alamos National Laboratory, Los Alamos, NM 87545}
\newcommand{\llnl}{Lawrence Livermore National Laboratory, Livermore, CA}
\newcommand{\lanla}{Los Alamos National Laboratory, Los Alamos, NM 87545}
\newcommand{\oxford}{Department of Physics, University of Oxford, 
Denys Wilkinson Building, Keble Road, Oxford OX1 3RH, UK}
\newcommand{\penn}{Department of Physics and Astronomy, University of 
Pennsylvania, Philadelphia, PA 19104-6396}
\newcommand{\pennx}{Department of Physics and Astronomy, University of 
Pennsylvania, Philadelphia, PA}
\newcommand{\queens}{Department of Physics, Queen's University, 
Kingston, Ontario K7L 3N6, Canada}
\newcommand{\uw}{Center for Experimental Nuclear Physics and Astrophysics, 
and Department of Physics, University of Washington, Seattle, WA 98195}
\newcommand{\uwx}{Center for Experimental Nuclear Physics and Astrophysics, 
and Department of Physics, University of Washington, Seattle, WA}
\newcommand{\uta}{Department of Physics, University of Texas at Austin, Austin, TX 78712-0264}
\newcommand{\triumf}{TRIUMF, 4004 Wesbrook Mall, Vancouver, BC V6T 2A3, Canada}
\newcommand{\ralimp}{Rutherford Appleton Laboratory, Chilton, Didcot, UK} 
\newcommand{\iusb}{Department of Physics and Astronomy, Indiana University, South Bend, IN}
\newcommand{\fnal}{Fermilab, Batavia, IL}
\newcommand{\uo}{Department of Physics and Astronomy, University of Oregon, Eugene, OR}
\newcommand{\hu}{Department of Physics, Hiroshima University, Hiroshima, Japan}
\newcommand{\slac}{Stanford Linear Accelerator Center, Menlo Park, CA}
\newcommand{\mac}{Department of Physics, McMaster University, Hamilton, ON}
\newcommand{\doe}{US Department of Energy, Germantown, MD}
\newcommand{\lund}{Department of Physics, Lund University, Lund, Sweden}
\newcommand{\mpi}{Max-Planck-Institut for Nuclear Physics, Heidelberg, Germany}
\newcommand{\uom}{Ren\'{e} J.A. L\'{e}vesque Laboratory, Universit\'{e} de Montr\'{e}al, Montreal, PQ}
\newcommand{\cwru}{Department of Physics, Case Western Reserve University, Cleveland, OH}
\newcommand{\pnnl}{Pacific Northwest National Laboratory, Richland, WA}
\newcommand{\uc}{Department of Physics, University of Chicago, Chicago, IL}
\newcommand{\mitt}{Laboratory for Nuclear Science, Massachusetts Institute of Technology, Cambridge, MA 02139}
\newcommand{\ucsd}{Department of Physics, University of California at San Diego, La Jolla, CA }
\newcommand{	\lsu	}{Department of Physics and Astronomy, Louisiana State University, Baton Rouge, LA 70803}
\newcommand{\imp}{Imperial College, London, UK}
\newcommand{\uci}{Department of Physics, University of California, Irvine, CA 92717}
\newcommand{\ucia}{Department of Physics, University of California, Irvine, CA}
\newcommand{\suss}{Department of Physics and Astronomy, University of Sussex, Brighton  BN1 9QH, UK}
\newcommand{\sussx}{Department of Physics and Astronomy, University of Sussex, Brighton, UK}
\newcommand{\lifep}{Laborat\'{o}rio de Instrumenta\c{c}\~{a}o e F\'{\i}sica Experimental de
Part\'{\i}culas, Av. Elias Garcia 14, 1$^{\circ}$, 1000-149 Lisboa, Portugal}
\newcommand{\lipx}{Laborat\'{o}rio de Instrumenta\c{c}\~{a}o e F\'{\i}sica Experimental de
Part\'{\i}culas,  Lisboa, Portugal}
\newcommand{\hku}{Department of Physics, The University of Hong Kong, Hong Kong.}
\newcommand{\aecl}{Atomic Energy of Canada, Limited, Chalk River Laboratories, Chalk River, ON K0J 1J0, Canada}
\newcommand{\nrc}{National Research Council of Canada, Ottawa, ON K1A 0R6, Canada}
\newcommand{\princeton}{Department of Physics, Princeton University, Princeton, NJ 08544}
\newcommand{\birkbeck}{Birkbeck College, University of London, Malet Road, London WC1E 7HX, UK}
\newcommand{\snoi}{SNOLAB, Lively, ON P3Y 1N2, Canada}
\newcommand{\snoix}{SNOLAB, Lively,  ON, Canada}
\newcommand{\uba}{University of Buenos Aires, Argentina}
\newcommand{\hvd}{Department of Physics, Harvard University, Cambridge, MA}
\newcommand{\pny}{Goldman Sachs, 85 Broad Street, New York, NY}
\newcommand{\pnv}{Remote Sensing Lab, PO Box 98521, Las Vegas, NV 89193}
\newcommand{\nts}{Nevada National Security Site, Las Vegas, NV}
\newcommand{\psis}{Paul Schiffer Institute, Villigen, Switzerland}
\newcommand{\liverpool}{Department of Physics, University of Liverpool, Liverpool, UK}
\newcommand{\uto}{Department of Physics, University of Toronto, Toronto, ON, Canada}
\newcommand{\uwisc}{Department of Physics, University of Wisconsin, Madison, WI}
\newcommand{\psu}{Department of Physics, Pennsylvania State University,
     University Park, PA}
\newcommand{\anl}{Deparment of Mathematics and Computer Science, Argonne
     National Laboratory, Lemont, IL}
\newcommand{\cornell}{Department of Physics, Cornell University, Ithaca, NY}
\newcommand{\tufts}{Department of Physics and Astronomy, Tufts University, Medford, MA}
\newcommand{\ucd}{Department of Physics, University of California, Davis, CA}
\newcommand{\unc}{Department of Physics, University of North Carolina, Chapel Hill, NC}
\newcommand{\dresden}{Institut f\"{u}r Kern- und Teilchenphysik, Technische Universit\"{a}t Dresden,  Dresden, Germany} 
\newcommand{\isu}{Department of Physics, Idaho State University, Pocatello, ID}
\newcommand{\qmul}{School of Physics and Astronomy, Queen Mary University of London, UK}
\newcommand{\ucsb}{Dept. of Physics, University of California, Santa Barbara, CA}
\newcommand{\cern}{CERN, Geneva, Switzerland}
\newcommand{\utah}{Dept. of Physics, University of Utah, Salt Lake City, UT}
\newcommand{\casa}{Center for Astrophysics and Space Astronomy, University
of Colorado, Boulder, CO}
\newcommand{\susel}{Sanford Underground Research Laboratory, Lead, SD}  
\newcommand{\ntu}{Center of Cosmology and Particle Astrophysics, National Taiwan University, Taiwan}
\newcommand{\berlin}{Institute for Space Sciences, Freie Universit\"{a}t Berlin,
Leibniz-Institute of Freshwater Ecology and Inland Fisheries, Germany}
\newcommand{\bhsu}{Black Hills State University, Spearfish, SD} 
\newcommand{\queensa}{Dept.\,of Physics, Queen's University, 
Kingston, Ontario, Canada} 
\newcommand{\aasu}{Dept.\,of Chemistry and Physics, Armstrong  State University, Savannah, GA}
\newcommand{\ucb}{Physics Department, University of California at Berkeley, Berkeley, CA 94720-7300}
\newcommand{\ucbx}{Physics Department, University of California at Berkeley, and Lawrence Berkeley National Laboratory, Berkeley, CA}
\newcommand{\mcgill}{Physics Department, McGill University, Montreal, QC, Canada}
\newcommand{\columbia}{Columbia University, New York, NY}
\newcommand{\rhul}{Dept. of Physics, Royal Holloway University of London, Egham, Surrey, UK}
\newcommand{\ubama}{Department of Physics and Astronomy, University of Alabama, Tuscaloosa, AL}
\newcommand{\kit}{Institut f\"{u}r Experimentelle Kernphysik, Karlsruher Institut f\"{u}r Technologie, Karlsruhe, Germany}
\newcommand{\winnipeg}{Department of Physics, University of Winnipeg, Winnipeg, Manitoba, Canada}
\newcommand{\kwantlen}{Kwantlen Polytechnic University, Surrey, BC, Canada}
\newcommand{\cea}{CEA-Saclay, DSM/IRFU/SPP, Gif-sur-Yvette, France}
\newcommand{\bankcanada}{National Bank of Canada, Montréal, QC}
\newcommand{\sunysb}{Laufer Center, Stony Brook University, Stony Brook, NY}
\newcommand{\rock}{Rock Creek Group, Washington, DC}
\newcommand{\rcnp}{Research Center for Nuclear Physics, Osaka, Japan}
\newcommand{\usd}{University of South Dakota, Vermillion, SD}
\newcommand{\lancaster}{Physics Department, Lancaster University, Lancaster, UK}
\newcommand{\potsdam}{GFZ German Research Centre for Geosciences, Potsdam, Germany}
\newcommand{\kirchhoff}{Ruprecht-Karls-Universit\"{a}t Heidelberg, Im Neuenheimer Feld 227, Heidelberg, Germany}
\newcommand{\continuum}{Continuum Analytics,  Austin, TX}
\newcommand{\gsu}{Dept. of Physics, Georgia Southern University, Statesboro, GA}
\newcommand{\pelmorex}{Pelmorex Corp., Oakville, ON} 
\newcommand{\usaid}{Global Development Lab, U.S. Agency for International Development, Washington DC}
    

\affiliation{\alta}
\affiliation{\ucb}
\affiliation{\ubc}
\affiliation{\bnl}
\affiliation{\carleton}
\affiliation{\uog}
\affiliation{\lu}
\affiliation{\lbnl}
\affiliation{\lifep}
\affiliation{\lanl}
\affiliation{\lsu}
\affiliation{\mitt}
\affiliation{\oxford}
\affiliation{\penn}
\affiliation{\queens}
\affiliation{\snoi}
\affiliation{\uta}
\affiliation{\triumf}
\affiliation{\uw}
\author{B.~Aharmim}\affiliation{\lu}
\author{S.\,N.~Ahmed}\affiliation{\queens}
\author{A.\,E.~Anthony}\altaffiliation{Present address: \usaid}\affiliation{\uta}
\author{N.~Barros}\altaffiliation{Present address: \pennx}\affiliation{\lifep}
\author{E.\,W.~Beier}\affiliation{\penn}
\author{A.~Bellerive}\affiliation{\carleton}
\author{B.~Beltran}\affiliation{\alta}
\author{M.~Bergevin}\altaffiliation{Present address: \llnl}\affiliation{\lbnl}\affiliation{\uog}
\author{S.\,D.~Biller}\affiliation{\oxford}
\author{R.~Bonventre}\affiliation{\ucb}\affiliation{\lbnl}
% ...
			
\collaboration{SNO Collaboration}
\noaffiliation

\date{\today}

\begin{abstract}
Neutron production in giga electron volt-scale neutrino interactions is a poorly studied process. We have measured the neutron multiplicities in atmospheric neutrino interactions in the Sudbury Neutrino Observatory experiment and compared them to the prediction of a Monte Carlo simulation using \textsc{genie} and a minimally modified version of \textsc{geant4}. We analyzed 837 days of exposure corresponding to Phase I, using pure heavy water, and Phase II, using a mixture of Cl in heavy water. Neutrons produced in atmospheric neutrino interactions were identified with an efficiency of $15.3\%$ and $44.3\%$, for Phases I and II respectively. The neutron production is measured as a function of the visible energy of the neutrino interaction and, for charged current quasielastic interaction candidates, also as a function of the neutrino energy. This study is also performed by classifying the complete sample into two pairs of event categories: charged current quasielastic and non charged current quasielastic, and $\nu_{\mu}$ and $\nu_e$. Results show good overall agreement between data and Monte Carlo for both phases, with some small tension with a statistical significance below $2\sigma$ for some intermediate energies.
\end{abstract}

\maketitle

\section{\label{sec:intro} Introduction}

During the past few years, great advances in our understanding of neutrino interactions in the \SI{100}{\MeV}$\sim$\SI{10}{\GeV} energy range have been achieved. Experiments like T2K\cite{t2k}, MiniBooNE\cite{miniboone}, and MINER$\nu$A\cite{minerva} have shed light on the neutrino-nucleus interaction mechanisms. Nevertheless, the limited ability of the detectors used by these experiments to identify the neutrons produced in the neutrino interactions limits our understanding of the interaction processes. Development of neutron tagging techniques is useful for three main reasons. First, it would reduce atmospheric neutrino backgrounds in proton decay or supernova relic neutrino searches, boosting the sensitivity of current experiments. Second, it could help to distinguish neutrinos from antineutrinos in nonmagnetized detectors, since antineutrinos typically produce more neutrons. Third, it would provide crucial information on neutrino cross section models, which are the driving systematic uncertainty in neutrino oscillation experiments like T2K and NO$\nu$A\cite{nova} and the future DUNE\cite{dune} and Hyper-Kamiokande\cite{hyperk}.

Water Cherenkov detectors have been proven to be of great value for solar and atmospheric neutrinos detection. Nevertheless, identification of neutrons generated in neutrino-nucleus interactions is challenging since it requires detection of the mega electron volt-scale deexcitation process that follows the neutron capture. Super-Kamiokande (SK) demonstrated that neutron detection is possible in water Cherenkov detectors \cite{sk_firstndet}, with a detection efficiency of approximately \SI{20}{\%}. In a later study, SK applied the new ability to measure the total number of generated neutrons in atmospheric neutrino interactions, as a function of the visible energy \cite{superk_neutrons}. However, no comparison between interaction models and measurements is provided, and such a comparison does not currently exist in the literature. In addition, an inclusive analysis is performed, without distinction between different types of neutrino-nucleus interactions.

In this study, neutrons produced in atmospheric neutrino interactions are successfully identified with the Sudbury Neutrino Observatory (SNO), a heavy water Cherenkov detector. A measurement of the number of produced neutrons as a function of visible energy of the neutrino interaction for different neutrino interaction types is presented along with a comparison with a Monte Carlo (MC) model using \textsc{genie} \cite{genie,genie_rw} and \textsc{geant4} \cite{geant4}. The number of produced neutrons as a function of reconstructed neutrino energy for charged current quasielastic events is also given. Finally, we study the potential for $\nu$ and $\bar{\nu}$ separation using neutron tagging.

This article is structured in the following way. A brief overview of the SNO detector is given in \Cref{sec:sno}, followed by a description of the MC model used in this analysis and a MC study in \Cref{sec:nprod}. The reconstruction algorithms used to characterize the atmospheric neutrino interactions and neutron captures are explained in \Cref{sec:reco}. The selection criteria for neutrino interactions and neutron captures are in \Cref{sec:vsel} and \Cref{sec:nsel}, respectively. \Cref{sec:syst} is dedicated to systematic uncertainties. The final measurements of neutron production in atmospheric neutrino interactions are presented in \Cref{sec:res}, along with a comparison to results from SK. \Cref{sec:discussion} presents the final discussion and summary.

\section{\label{sec:sno} SNO detector}

\begin{figure}[!ht]
	\includegraphics[width=\columnwidth]{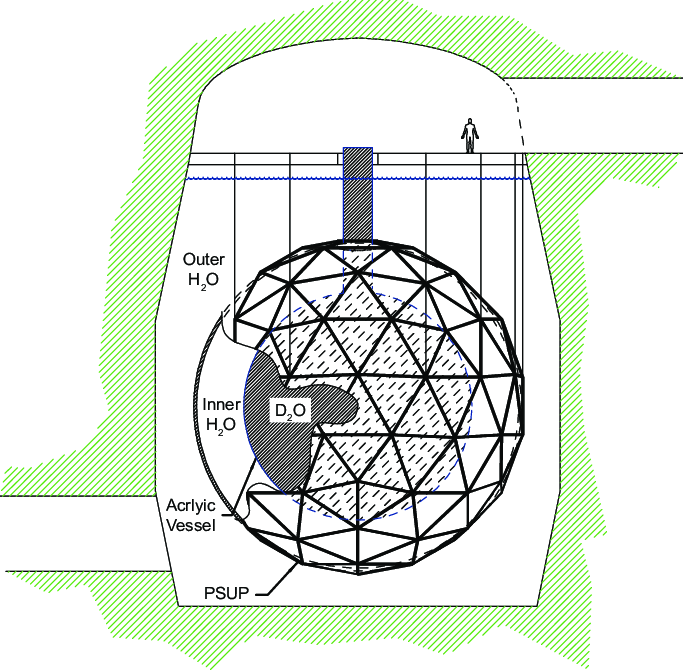}
	\caption{\label{fig:sno} The SNO detector. The labels correspond to the different volumes.}
\end{figure}

SNO was a Cherenkov detector using heavy water located at a depth of \SI{2092}{\m} ($5890$~mwe) in INCO's Creighton mine, near Sudbury, Ontario. The layout of the detector is shown in \Cref{fig:sno} and it consisted of a \SI{6}{\m} radius spherical acrylic vessel (AV) volume containing heavy water nested into an \SI{8.4}{\m} radius spherical structure instrumented with 9456 photomultipliers (PMTs) \cite{sno_det}. The total mass of the detector enclosed in the PMT structure, adding the heavy and light water regions, was about \SI{2.7}{kt}. The entire detector was suspended in a cavity and submerged in light water, which shielded against radioactivity from the rock. A cylindrical tube called the neck connected the inner part of the acrylic vessel with an external clean room, which served as the interface for filling and deploying calibration sources. The outer detector region featured 91 PMTs attached to the main structure but facing outward (referred as OWL), in order to provide a veto against external events. In addition, 8 PMTs (referred as NECK) were attached inside the neck, and 23 PMTs were suspended in a rectangular frame in the outer light water volume facing towards the neck region. The motivation was to veto possible light leaks occurring at the interface of the detector with the deck, and the flashes of light that were produced at the interface between the acrylic and the water surface.

The SNO experiment was designed for solar neutrino detection, and hence it was optimized for low-energy events. Neutron captures on heavy water provide a higher-energy signal than conventional water Cherenkov detectors. This increases their observable energy above the typical radioactive backgrounds, and allows a higher neutron detection efficiency. SNO was operated in three different phases. In Phase I (the \DTO phase), the active volume was filled with pure heavy water. In Phase II (the salt phase), the heavy water volume was doped with chlorine in salt form (NaCl) at \SI{0.2}{\%} by weight, which considerably boosted the neutron capture cross section and signal energy. Finally, in Phase III, $^3$He proportional counters were deployed in the detector, which provided a completely independent means of neutron detection. However, this last phase is not used in the current analysis due to the added complexity to the geometry, which would require further study to determine the impact on our reconstruction of atmospheric neutrino interactions. The results reported in this analysis correspond to data collected during $337.25\pm0.02$ days for Phase I and $499.45\pm0.02$ days for Phase II.

\section{\label{sec:nprod} Neutron production and detection in atmospheric neutrino interactions}

Production of neutrons in neutrino interactions is a complicated process that depends on neutrino-nucleon cross sections; on the interactions of the produced particles within the nuclear media, known as the final state interactions (FSIs); and on the hadronic interactions of the final state particles that propagate in the detector media. We differentiate two ways neutrons can be produced in atmospheric neutrino interactions:
\begin{enumerate}
	\item In the final state of the neutrino-nucleus interaction (\textit{primary neutrons}): this includes neutrons produced directly at the interaction vertex by antineutrinos, as well as those created due to FSIs.
	\item As the byproduct of interactions of final state particles in the detector media (\textit{secondary neutrons}): this includes neutron production due to hadronic inelastic scattering, photonuclear interactions of leptons and mesons, and muon captures.
\end{enumerate}

The free neutrons propagate in the detector media undergoing nuclear collisions before they are captured. Since the energy of the produced neutrons is much higher than \SI{1}{\keV} (fast neutrons), they need to reach thermal energies (approximately \SI{0.025}{\eV}) prior to being captured. The number of scatters they undergo strongly depends on the neutron energy. In heavy water, the energy loss is on average $44\%$ per collision, so the number of scatters for neutrons between \SI{1}{\MeV} and \SI{1}{\GeV} can range between 10 and 30, with higher-energy neutrons being more likely to exit the detector. Following the thermalization process, the neutron diffuses in the medium until it is captured. This diffusion is orders of magnitude slower than thermalization, so the neutron capture time is mostly determined by diffusion, which is specific to the capture material and independent of the energy at which the neutron was produced. Finally, the neutron is captured by a nucleus, which is left in an excited state and will deexcite, emitting particles on a very short timescale. The processes that could lead to a significant neutron detection in SNO are neutron captures on H, $^2$H, $^{35}$Cl, and $^{16}$O, with a subsequent emission of gamma rays of energies \SI{2.2}{\MeV}, \SI{6.25}{\MeV}, a cascade of \SI{8.6}{\MeV}, and a cascade of \SI{4.1}{\MeV}, respectively. Since the \SI{2.2}{\MeV} gamma-ray from H capture is below our analysis energy threshold, this detection channel is not relevant.

The entire process from neutron production to capture is simulated by our MC model. \textsc{genie} is used as a neutrino interaction generator, producing the final state particles, including primary neutrons. These particles are further propagated in the SNO geometry using \textsc{geant4}, which handles generation of secondary neutrons, neutron transport, capture, and gamma-ray emission. Finally, the detection process of gamma rays is handled by the \textsc{snoman}\cite{sno_det} detector simulation, which models the detector response. In the following section, we detail each stage of the simulation.

\subsection{\label{sec:genie} Generating neutrino interactions with GENIE}

Atmospheric neutrinos interact in the different volumes of the SNO detector through charged current (CC) and neutral current (NC) interactions. Since the neutrino energies span several orders of magnitude, neutrinos will undergo several types of interactions: elastic scattering (ES), quasielastic (QE), resonant pion production (RES), deep inelastic scattering (DIS), or coherent scattering (COH). Pions and other hadrons will undergo a variety of FSI processes, such as: pion absorption, charge exchange, pion production, and elastic scattering, that modify the kinematics and nature of the original particles.

The neutrino interaction generator \textsc{genie} (version 2.10.2) is used to generate atmospheric neutrino interactions, the complex interaction models of which are described in Ref. \cite{genie}, and the most relevant parameters for our analysis are summarized in \Cref{tab:syst_xsecs}. We input the unoscillated Bartol04 neutrino flux calculated for the SNOLAB location \cite{bartol04} and the SNO geometry and material composition for each phase. Neutrino oscillations are treated subsequently by reweighing the events. The total simulated data set contains 2 orders of magnitude more events than expected for the exposure of the analyzed data.

\subsection{\label{sec:geant4} Secondary neutron generation and neutron propagation in GEANT4}

The final state particles produced by \textsc{genie} are used as input into the \textsc{geant4} tool-kit (version 10.0) \cite{geant4}, using the shielding physics list version 2.1. The same detector geometry used for \textsc{genie} is used in this step. The generation of neutrons is handled by a number of different models that simulate the processes: gamma photonuclear interactions; muon and electron nuclear processes; and inelastic scattering of mesons, protons and neutrons. Some of these processes have been compared against model predictions \cite{n_geant4_fluka,n_geant4_proton}. A limitation of GEANT4 is that it does not properly simulate deuteron photonuclear breakup. The impact of this process was estimated to be below \SI{0.4}{\%} by using an implementation of the original model developed for the SNO experiment \cite{sno_det}. Neutron elastic scattering, crucial for the simulation of the thermalization process, is modeled using the \textsc{NeutronHP} package for energies below \SI{20}{\MeV} and the \textsc{chips} model for the higher energy range \cite{geant4}. This is a data-driven model that uses the Evaluated Nuclear Data File database. The relevant processes for neutron capture are also implemented in \textsc{NeutronHP}. A known problem with this model is that it randomizes the energy of the emitted gamma rays. As a result the sum of the total energy does not correspond to the actual total energy available for the deexcitation, violating energy conservation. This is not an issue for $^2$H and $^3$H, where a single energy state is present, but it is incorrect for $^{17}$O and $^{36}$Cl. A custom model based on the SNO implementation had to be introduced. We created a new neutron capture final state in our local \textsc{geant4} installation that includes the actual branching ratios for $^{17}$O deexcitations and for $^{36}$Cl. The used energy levels and branching ratios for $^{36}$Cl are extracted from Ref. \cite{ncap_cl36}.

\begin{table}[bh!]
	\begin{tabular}{c|c}
				Neutron origin & Fraction \\
				\hline
				\hline
				Neutrino interaction & $33.0 (0.2)\%$ \\
				\hline
				Neutron inelastic & $34.9 (0.2)\%$ \\
				$\pi$/$K$ inelastic & $15.0 (0.1)\%$ \\
				Proton inelastic & $7.3 (0.1)\%$ \\
				Hadron capture at rest & $6.4 (0.1)\%$ \\
				$\mu$ capture at rest & $2.20 (0.04)\%$ \\
				Photonuclear & $0.90 (0.02)\%$ \\
				Other & $0.29 (0.01)\%$ \\
			\end{tabular}
	\caption{\label{tab:neutron_origin} Origin of neutrons produced by atmospheric neutrinos in SNO as predicted by the MC simulation. The processes below the single horizontal line correspond to the sources of secondary neutrons. The uncertainties in parentheses correspond to the MC statistical uncertainties.}
\end{table}

\begin{figure}[th!]
	\centering
	\includegraphics[width=\columnwidth]{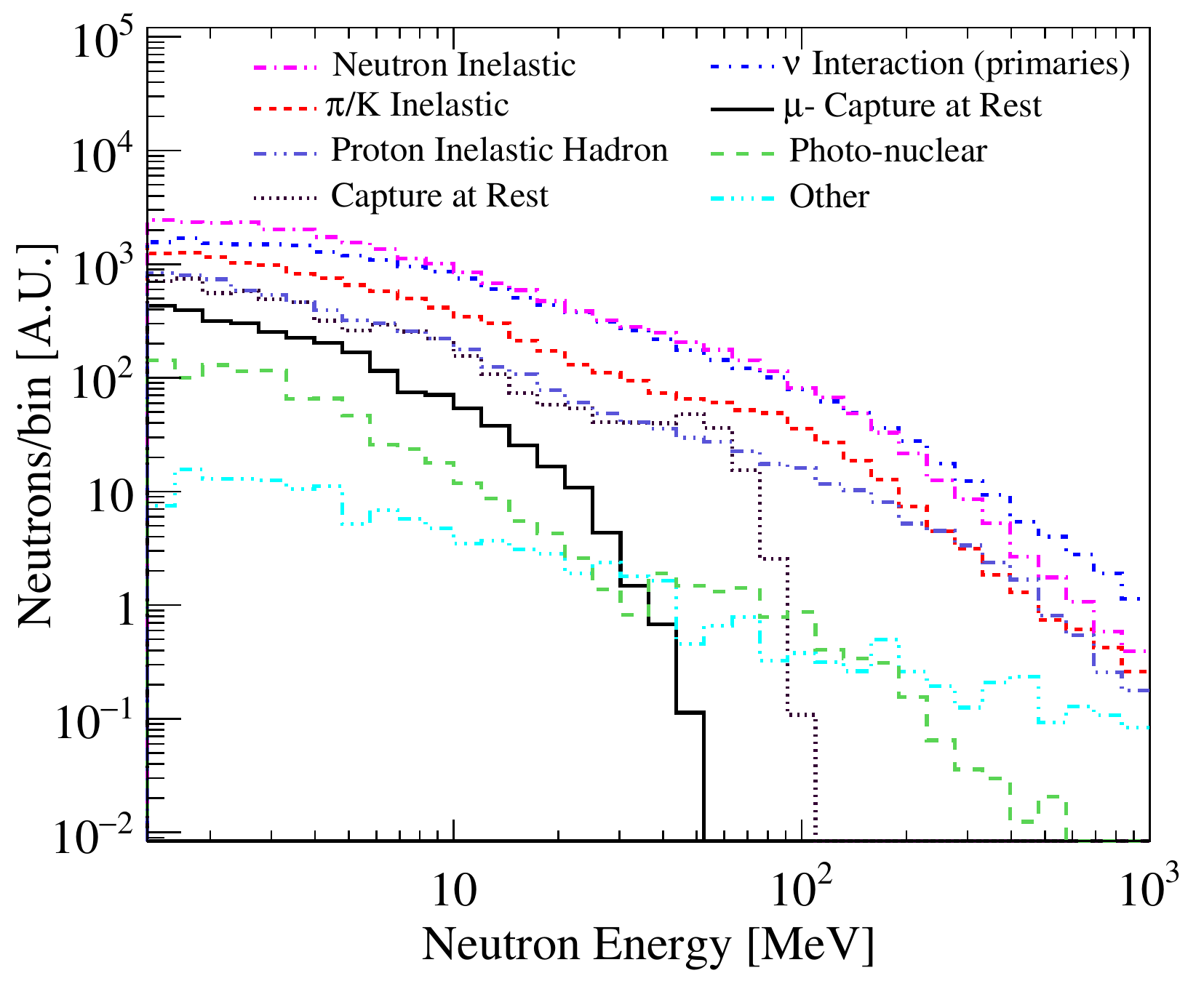}
	\caption{\label{fig:neutron_ke} Neutron kinetic energy distributions broken down by neutron origin, as predicted by the MC simulation.}
\end{figure}

\begin{figure*}
	\centering
	\includegraphics[width=0.75\textwidth]{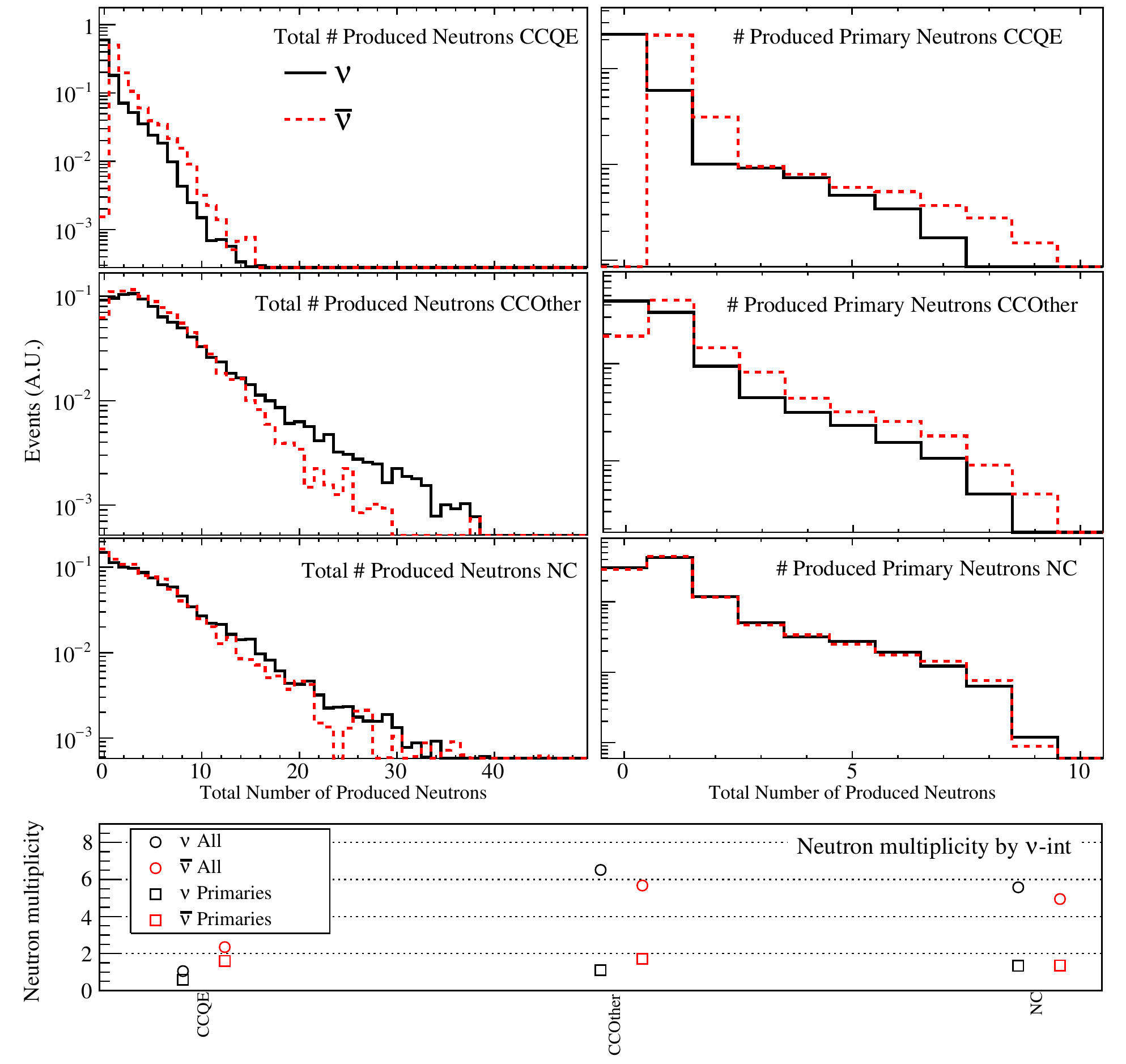}
	\caption{\label{fig:nprod_int} Predicted neutron production in the SNO detector per event for different neutrino interactions (rows) with no event selection applied. Total neutron production is shown on the left, and only primary neutrons are shown on the right. At the bottom, the average number of neutrons is shown for each case.}
\end{figure*}

\begin{figure*}
	\centering
	\includegraphics[width=0.75\textwidth]{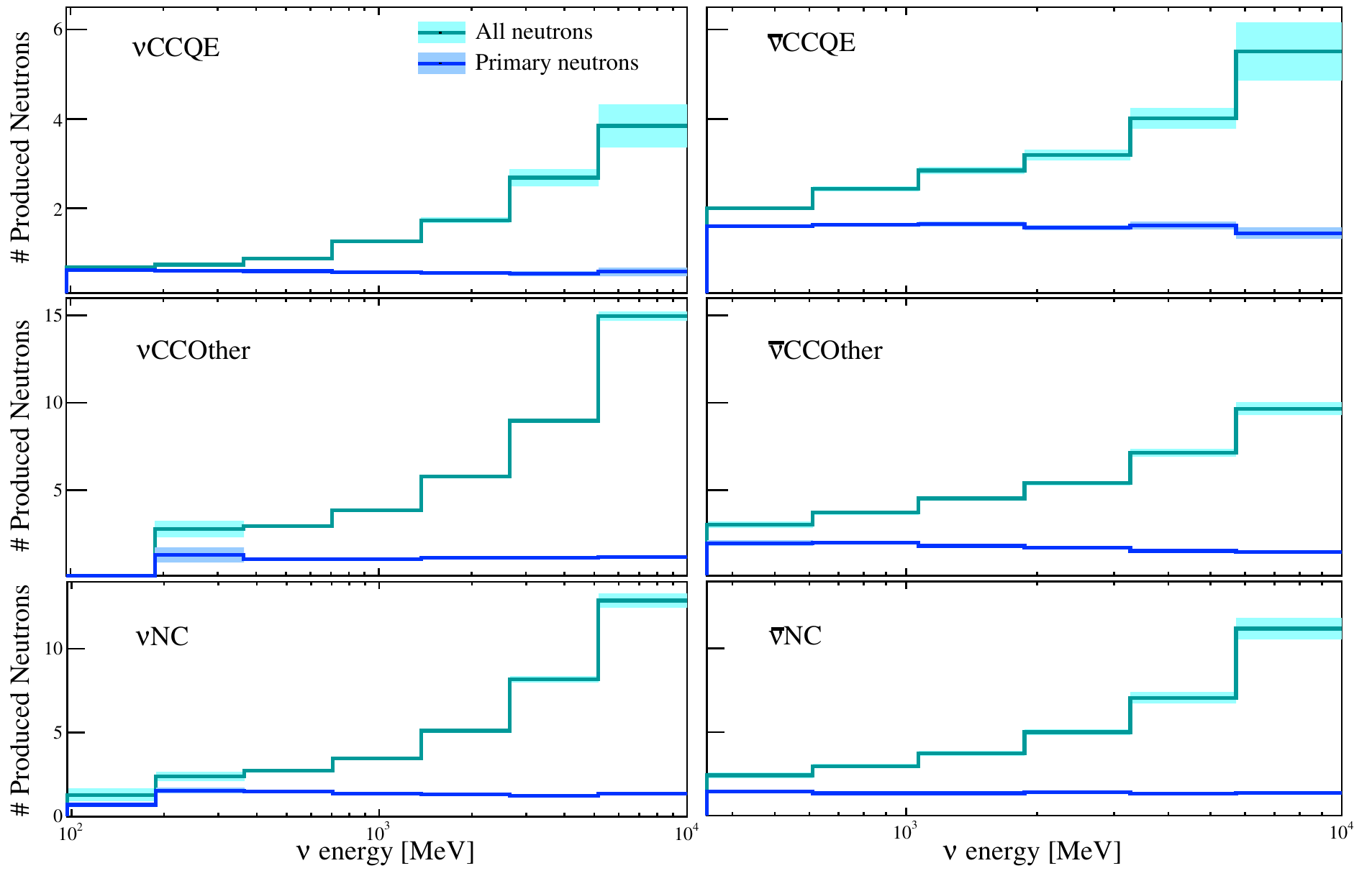}
	\caption{\label{fig:nprod_venergy} Predicted neutron production in the SNO detector as a function of the neutrino energy for different neutrino interactions and for neutrinos (left) and antineutrinos (right). No event selection is applied.}
\end{figure*}

A breakdown of the origin of the neutrons produced along with their energy distributions is shown in \Cref{tab:neutron_origin} and \Cref{fig:neutron_ke}, where we observe that roughly one-third of the neutrons is primary neutrons; one-third is produced as a result of neutron scattering, and one-third is due to other processes involving mainly protons, mesons and leptons. The energy of the produced neutrons ranges from a few mega-electron-volts to \SI{1}{\GeV}, approximately $90\%$ of them being below \SI{50}{\MeV}. The total number of produced neutrons in CCQE interactions, other CC interactions (CCOther) and NC interactions for neutrinos and antineutrinos is shown in \Cref{fig:nprod_int}. We observe that $69.5(0.8)\%$ of the neutrino interactions produces at least one neutron, as summarized in \Cref{tab:nprod_number}. On average, antineutrinos produce approximately one more primary neutron than do neutrinos in CC interactions, as can be seen at the bottom of \Cref{fig:nprod_int}. This difference is washed out by the production of secondary neutrons in CCOther interactions, but it still holds for CCQE interactions, highlighting the potential for $\nu$-$\bar{\nu}$ separation. The production of secondary neutrons is similar to the production of primary neutrons in CCQE interactions, but this is much larger in CCOther and NC interactions. The neutron production as a function of neutrino energy is shown in \Cref{fig:nprod_venergy}. Although the charged hadron production increases with the invariant hadronic mass, and hence neutrino energy \cite{genie_hadronic}, the production of primary neutrons is practically constant over the entire energy range, and it is only the production of secondary neutrons that leads to an increase of the overall neutron multiplicity. According to our MC model, the fraction of neutrons that are produced within the AV and also captured inside the AV is $31.1\pm 0.3\%$ for Phase I and $74.4\pm 0.4\%$ for Phase II.

\begin{table}[!h]
	\centering
	\begin{tabular}{c|c}
		\multirow{2}{*}{Process}	& Fraction with at least \\
		& one neutron produced \\
		\hline
		$\nu$ CCQE  & 38.4(2.2)\% \\
		$\bar{\nu}$ CCQE & 99.9(0.1)\% \\
		$\nu$ CCOther & 88.8(2.0)\% \\
		$\bar{\nu}$ CCOther & 94.7(2.1)\% \\
		$\nu$ NC & 84.8(1.8)\% \\
		$\bar{\nu}$ NC & 82.4(2.3)\% \\
		\hline
		$\nu$ total & 61.5(1.1)\% \\
		$\bar{\nu}$ total & 95.6(0.6)\% \\
		Total & 69.5(0.8)\%
	\end{tabular}
	\caption{\label{tab:nprod_number} Percentage of events producing at least one neutron. The calculated uncertainties in parentheses corresponds to the MC statistical uncertainty.}
\end{table}

\subsection{\label{sec:cali} Detector simulation}

The SNO detector is simulated with the package developed for the original SNO analyses, \textsc{snoman} \cite{sno_det}. This package handles production and propagation of Cherenkov light in realistic detector conditions. The status of the electronics was recorded and simulated on a run by run basis, including the number of working PMTs and trigger conditions. Then, run-dependent efficiencies or reconstruction biases were modeled by \textsc{snoman}, which was extensively calibrated and validated using different deployed sources including AmBe and $^{252}$Cf to study the neutron detection response, $^{16}$N to calibrate the energy scale, and a diffused laser source to measure the optical properties of the detector \cite{sno_det}. We also use \textsc{snoman} to simulate Cherenkov production from the final state particles produced by \textsc{genie} and \textsc{geant4}.

\section{\label{sec:reco} Event reconstruction}

Two different classes of events need to be characterized: atmospheric neutrino interactions, which produce high-energy (approximately giga-electron-volt) leptons and hadrons in the final state with well-defined ringlike Cherenkov images in the detector, and neutron captures, which produce lower-energy (approximately mega-electron-volt) gamma rays that give a less well-defined Cherenkov signal. In order to properly deal with these different energy ranges, two event reconstruction algorithms are used and described below.

\subsection{\label{sec:rf} Reconstruction of atmospheric neutrino interactions}

The atmospheric neutrino reconstruction algorithm called \textsc{Ring Fitter} \cite{richie_thesis} is designed to provide the position, direction, energy, particle identification (PID), and particle multiplicity from an atmospheric neutrino interaction occurring in the detector. The final state charged particles from a neutrino interaction are typically above approximately \SI{50}{\MeV}, so the directional nature of the Cherenkov light creates well-defined ringlike structures. Characterizing these rings gives us critical information on the nature of the particle and consequently the neutrino interaction. The algorithm is based on the routines used by Cherenkov detectors such as MiniBooNE \cite{miniboone_fitter} and Super-Kamiokande \cite{sk_fitter}. In the following, we give an overview of the algorithm.

\subsubsection{Preliminary ring identification}
We use the Hough transform technique \cite{hough} to identify the center of the main ring in the spherical surface defined by the PMT structure. This will serve to give a preliminary estimate of the particle direction.

In order to obtain a first estimate of the event position, the fitter developed for the SNO+\cite{snop_nd} water phase is used. Since it is optimized for low-energy events by design, its performance is poor at giga-electron-volt energies and it does not provide information on the particle type or multiplicity. The obtained position is used as a seed for the subsequent more complex algorithm.

The particle energy is also estimated at this stage by using the preliminary event position and the total amount of light collected in the event. This is done by building a lookup table using a complete MC simulation. Electrons and muons of energies up to \SI{2}{\GeV} and at different positions in the detector are generated using \textsc{snoman}. The result is a map of position and total charge vs energy.

\subsubsection{Determination of event position and direction}
A likelihood fit is performed under the single-ring hypothesis to find the following observables related to the highest-energy particle, referred to as the \textit{main ring}: event position $\vec{r}$, event time within the event window $t_e$ and event direction $\vec{d}$. The fit is run twice, once assuming an electron and again assuming a muon. The value of the likelihood in each case helps in identifying particle type, as described in the next section. The likelihood fit is based on the prediction of the number of photoelectrons (p.e.) that would be produced in each PMT for a specific position, direction, energy and particle hypothesis, represented by $\vec{x} = (\vec{r}, \vec{d}, t_e)$. The probability of observing $n$ p.e. in a single PMT when $\lambda$ p.e. are expected is assumed to follow a Poisson distribution:
\begin{equation}
	P_N(n|\lambda) = \frac{e^{-\lambda}}{n!}\lambda^n.
\end{equation}

For a given $n$, each PMT hit would present a different time and charge distribution, depending on its position with respect to the Cherenkov cone. The PMT time residual is defined as the PMT hit time corrected by the light's time of flight assuming a position for the emission of the photon, which corresponds to $\vec{r}$. The probability of observing a hit $i$ with charge $q_i$ and time residual $t_i(\vec{r})$ for a given $n_i$ and $\vec{x}$ hypothesis will be the product of the charge and time probabilities, $P_Q$ and $P_T$:
\begin{equation}
	P_Q(q_i|n_i) \times P_T(t_i(\vec{r})|n_i)
\end{equation}
which are defined below. Then, the probability that a PMT $i$ with $\lambda_i$ expected p.e. records a hit with a given $q_i$ and $t_i$ is obtained by summing over $n$:
\begin{eqnarray}
	P^{hit}_i(q_i,t_i(\vec{r})|\lambda_i) = \sum_{n_i} P_N(n_i|\lambda_i) \times P_Q(q_i|n_i) \nonumber \\
	\times P_T(t_i(\vec{r})|n_i)
\end{eqnarray}

If the $j$th PMT is not hit, then $n=0$ and the probability will simply be
\begin{equation}
	P^{unhit}_j = e^{-\lambda_j(\vec{x})}
\end{equation}

The likelihood function is obtained by multiplying the previous probabilities for all hit and unhit PMTs:
\begin{equation}
	\mathcal{L}(\vec{x}) = \prod^{hit}_i P^{hit}_i(q_i,t_i(\vec{r})|\lambda_i(\vec{x})) \prod^{unhit}_j e^{-\lambda_j(\vec{x})}
\end{equation}

For the PMT charge distribution we use the SNO single p.e. model and the averaged PMT gain measured at the detector. $P_Q(q_i|n_i)$ is generated from MC using the measured signal p.e. charge distribution. On the other hand, the time distribution for single p.e. is parametrized as a prompt and prepulse peak, plus a uniform noise contribution and a flat scattering contribution for $t>\SI{0}{\ns}$. This distribution will be skewed towards earlier times for multi-p.e. hits, since the time registered by a PMT corresponds to the earliest photon. To model this effect, we created a two-dimensional probability distribution function (PDF) of $P_T$ as a function of $n$. This is done by extracting $n$ times from the single p.e. time distribution and populating the new PDF taking the time of the earliest p.e.

Estimation of $\lambda$ is done differently for muons and electrons. Muons created by atmospheric neutrino interactions are typically minimum ionizing particles during most of their range and suffer very little scattering. These two features are important since as a result the energy loss, path, and Cherenkov production per unit length are very reproducible for every muon; they typically travel on fairly straight lines, yielding a well-defined Cherenkov cone with a thickness proportional to their energy. Then, the Cherenkov yield and topology are determined very well by the position where the muon is created, along with its direction and energy. To estimate $\lambda$, we use a MC-generated PDF as a function of the PMT angle and distance from the muon track. For electrons, since their paths are much shorter, we approximate them as points. The angular dependence of the number of produced p.e. is calculated using the MC simulation for different electron direction and energy hypotheses.

Finally, we find the best fit value by floating $\vec{x}$ and using the \textsc{minuit} routine implemented in \textsc{root}\cite{root}. We use the \textsc{migrad} algorithm to find the fit position and direction $\vec{x}_f$, for each of the two particle hypotheses.

\subsubsection{Particle identification and energy reconstruction}
We identify whether the particle is electronlike or muonlike by exploiting the fact that the angular distribution of the emitted photons is much broader for electrons than for muons, due to the more pronounced electron scattering and secondary gamma-ray emission. We run the likelihood fit described above under the electron and muon hypotheses and calculate the likelihood difference $\Delta\mathcal{L}$ to determine particle type. The hypothesis with the best fit value is taken as the particle type. In cases where the fit for the position $\vec{r}$ is poor, the difference between the two hypotheses becomes small and the particle identification degrades. To overcome this problem, the likelihood is recomputed without the time residual term $P_T(t_i(\vec{r})|n_i)$, and again, the hypothesis with the best fit value is chosen.

After the position, direction, and particle type have been precisely determined, we recalculate the particle energy by using MC-generated lookup tables for electrons and muons, binned in total PMT charge and radial position. The visible energy is defined as the electron-equivalent energy, i.e., the energy needed by an electron to produce the number of detected p.e. at the reconstructed radial position. The muon-equivalent energy is calculated in a similar fashion, and it is used to reconstruct the neutrino energy of muonlike events (see \Cref{sec:venergy}).

\subsubsection{Determination of ring multiplicity}
Once the first ring has been identified and characterized, we predict the number of p.e. for each PMT and subtract them from the event. Then, the Hough transform is computed again in order to look for secondary rings. The predicted total charge for the $i$th PMT is defined by the average charge for the estimated number of p.e. $\lambda_i$ given by
\begin{equation}
	\sum_{n_i>0} q_i \times P_N(n_i|\lambda_i) \times P_Q(q_i|n_i)
\end{equation}

In order to reject false secondary rings, a Kolmogorov-Smirnov (KS) test against a flat background is performed. The used distribution is that of the PMT charge as a function of the angle between the PMT positions and the reconstructed center of the ring. An event is tagged as multiring if the total absolute charge and charge densities are above a certain threshold computed from MC and if the KS value is not significant. Should any of these conditions fail, the event is considered to be single ring.

\subsubsection{Estimation of neutrino energy}
\label{sec:venergy}
\begin{figure}[!h]
	\centering
	\includegraphics[width=\columnwidth]{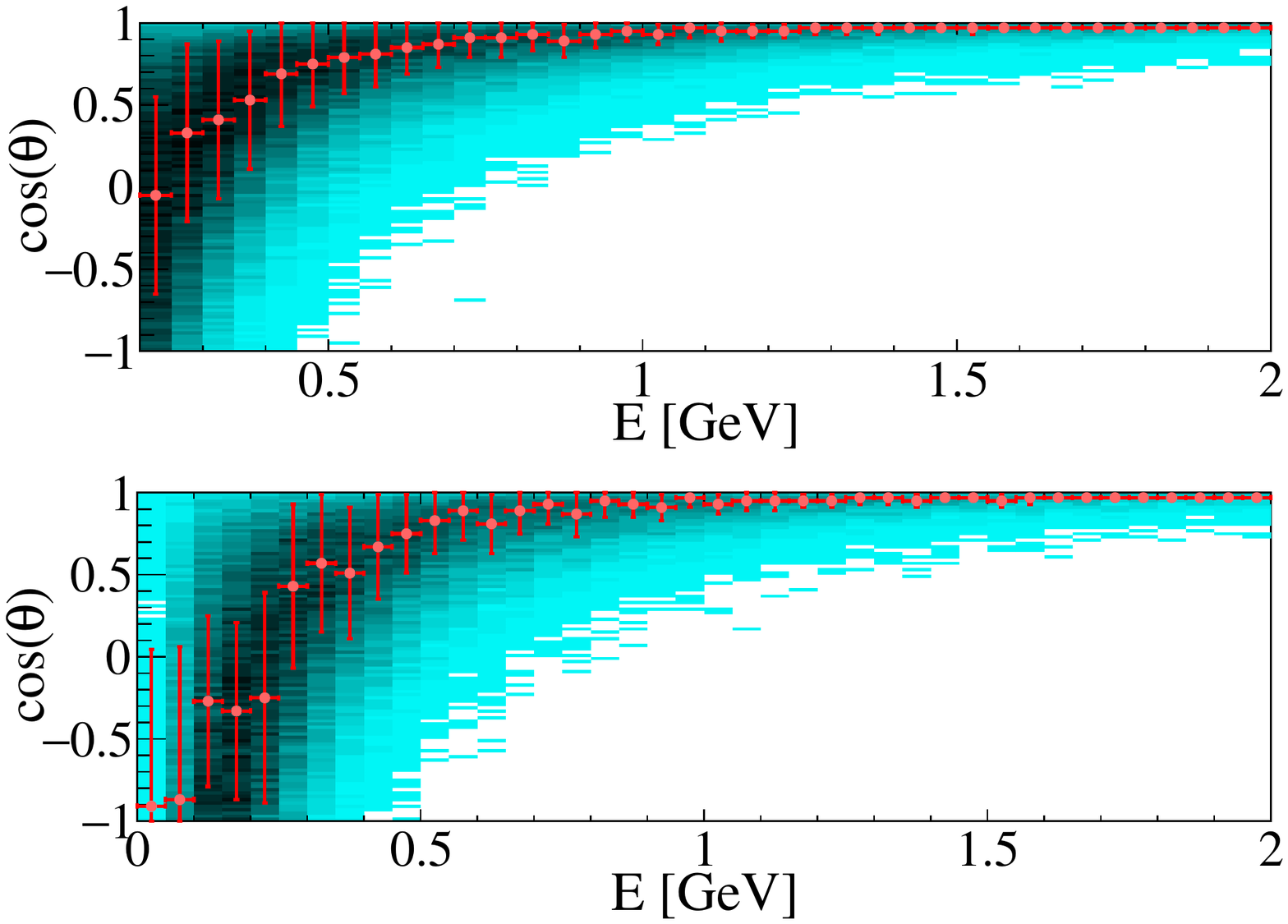}
	\caption{\label{fig:vreco_costh} Angle of produced lepton $\cos\theta_l$ vs lepton energy in a CCQE neutrino interaction for muons (top) and electrons (bottom). The red dots show the mode of the $\cos \theta$ distribution at each energy bin with the $1\sigma$ uncertainty.}
\end{figure}

The neutrino energy is reconstructed according to the CCQE hypothesis,
\begin{equation}
\label{eq:vreco} E^{\nu}_r = \frac{m_p^2 - (m_n - E_b)^2 - m_l^2 + 2(m_n-E_b)E_l}{2(m_n - E_b - E_l + p_l\cos\theta_l)}	
\end{equation}
where $m_p$, $m_n$, and $m_l$ are the masses of the proton, neutron, and charged lepton, $E_b=\SI{27}{\MeV}$ is the effective binding energy of a nucleon in oxygen for leptonic interactions \cite{E_b}, $E_l$ is the energy of the charged lepton, and $\cos\theta_l$ is the angle between the outgoing lepton and the incoming neutrino. Since the atmospheric neutrino direction is unknown, we estimate $\cos \theta_l$ from the \textsc{genie} prediction as the mode of the $\cos \theta_l$ distribution in a charged lepton's energy bin (see \Cref{fig:vreco_costh}). In this way, only the energy of the charged lepton is needed to estimate the neutrino energy. The uncertainties in these curves are computed by defining a symmetric region around the mode that encloses $68\%$ ($1\sigma$) of the events in each energy bin.

\subsection{Performance of reconstruction of atmospheric neutrino interactions}

The \textsc{Ring Fitter} algorithm has been validated against MC simulation of single particles and neutrino interactions. Single muons and electrons are generated across the detector volume at energies between 0 and \SI{2}{\GeV}. The energy resolution, position resolution, particle misidentification, and ring miscounting have been validated as a function of the energy and radius with electron and muon simulations. In the energy region of interest, the radial position resolution is \SI{28}{\cm} on average, the charged lepton energy resolution is below $7\%$, the particle misidentification rate is below $17\%$, and the rate of identification of single-ring events as multiring events is below $10\%$.

\begin{figure}[h!]
	\centering
	\subfigure[\label{fig:vbias_rad} Bias of reconstructed radial position with respect to the true position of the interaction for simulated atmospheric neutrino events.]{\includegraphics[width=0.96\columnwidth]{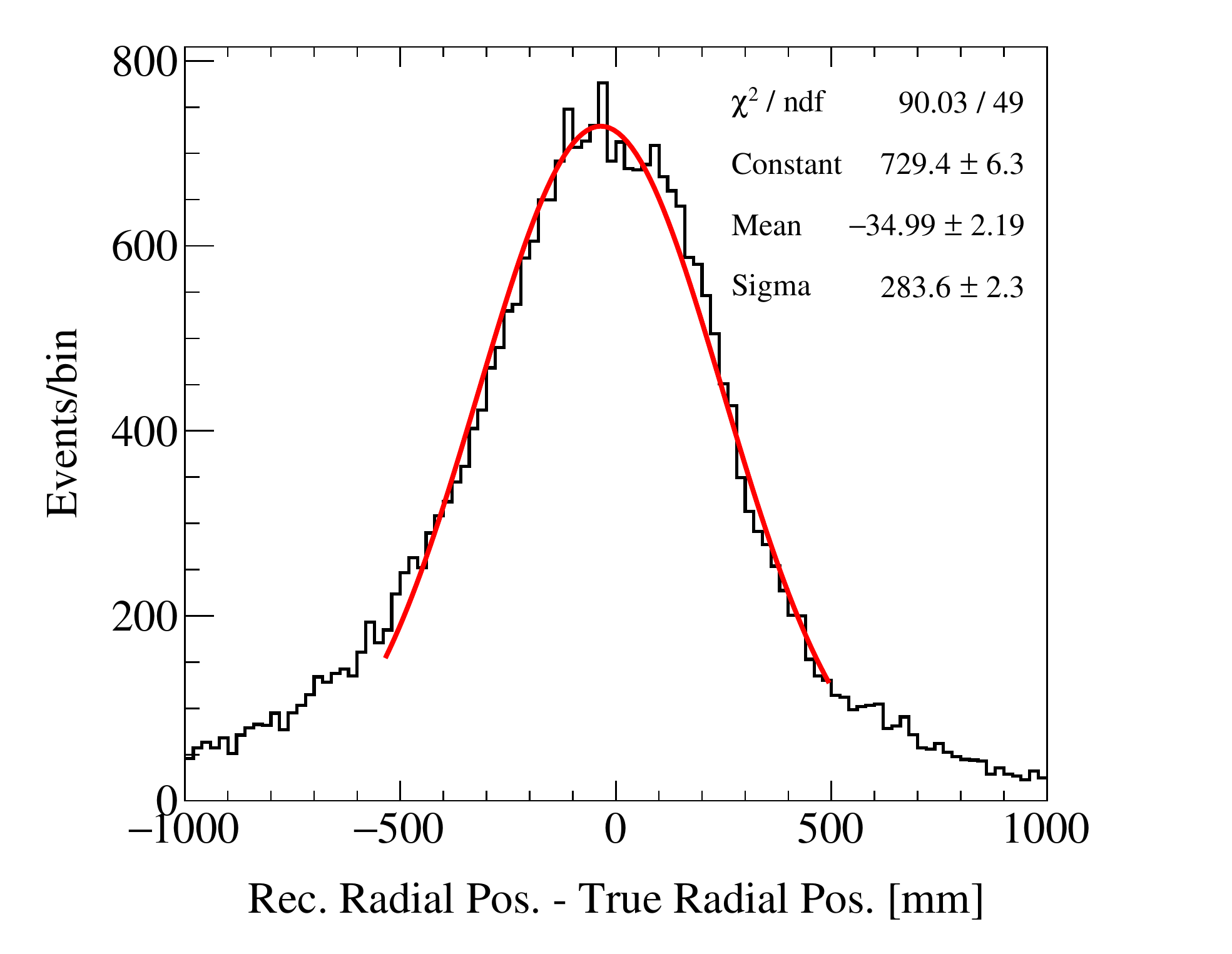}}
	\subfigure[\label{fig:vbias_vene} Reconstructed neutrino energy bias.]{\includegraphics[width=0.96\columnwidth]{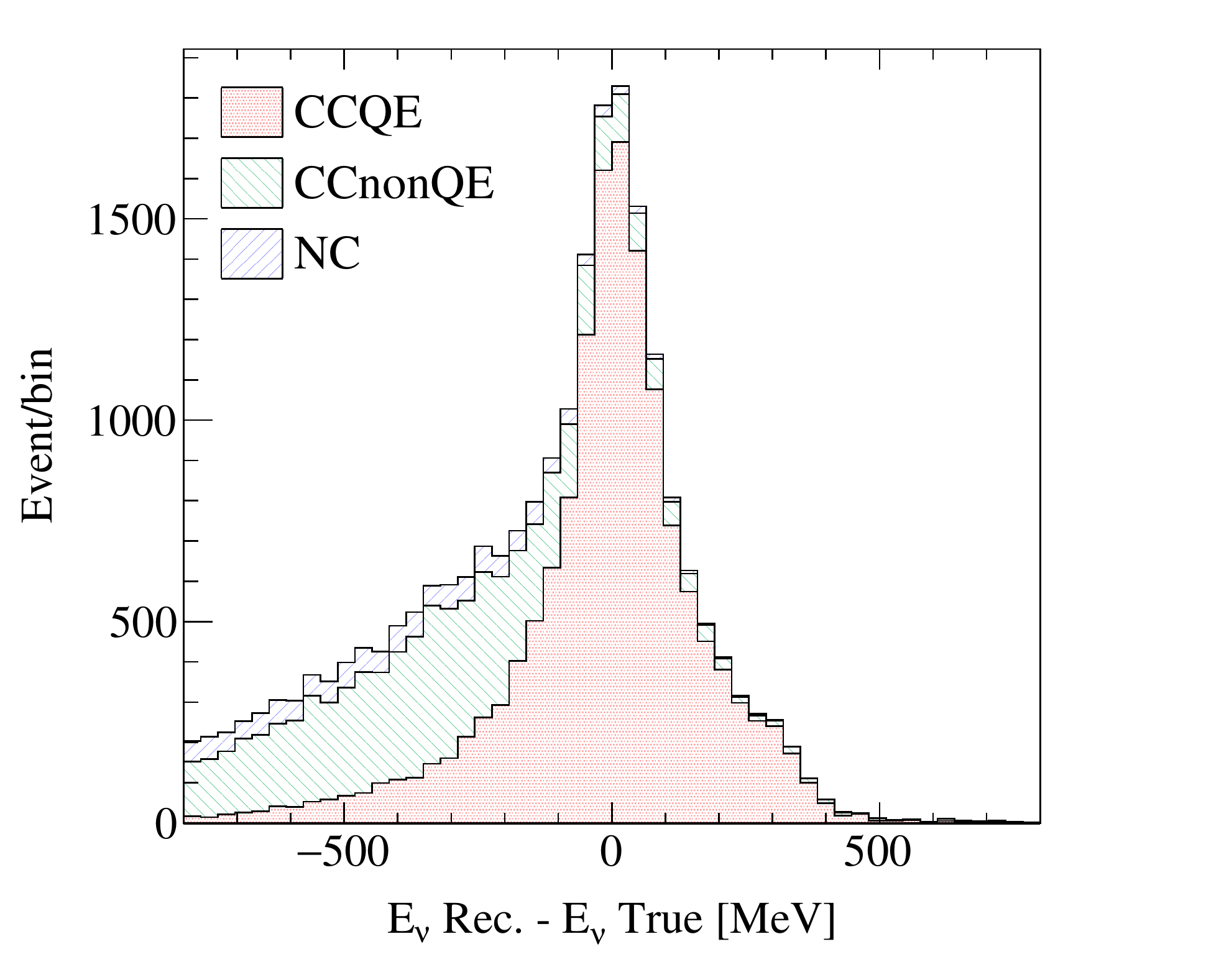}}
	\caption{Validation of reconstruction of simulated neutrino interactions.}
\end{figure}

The reconstruction of atmospheric neutrino interactions was validated using simulated events by comparing the reconstructed radial position and neutrino energy with the true values. The bias in the radial position is very small and below the position resolution, as shown in \Cref{fig:vbias_rad}. The bias in the reconstructed neutrino energy using the CCQE hypothesis is shown in \Cref{fig:vbias_vene}. The CCQE events have a negligible bias of \SI{7\pm1.2}{\MeV}, while the other type of interactions exhibit a significant deviation, as expected since they do not obey the CCQE hypothesis.

\subsection{\label{sec:lowreco} Reconstruction of neutron captures}

To extract information on neutron captures, the official SNO reconstruction algorithms are used, which have been extensively validated with calibration sources. The position is reconstructed using the so-called path fitter, and the energy is measured by the \textsc{ftk} algorithm, described in Ref. \cite{ftk}. These yield an approximately \SI{15}{\%} energy resolution and an approximately \SI{20}{\cm} position resolution for event energies of \SI{6}{\MeV}, estimated using an $^{16}$N source \cite{sno_n16}.

\section{\label{sec:vsel} Selection of atmospheric neutrino events}

Atmospheric neutrinos energies above \SI{40}{\MeV} are selected, so their interaction in the SNO detector produces charged particles well above the radioactive backgrounds. Atmospheric neutrino candidates are identified by criteria that start with the selection of events with more than 200 triggered PMTs (NHits). Additional cuts are designed to minimize instrumental backgrounds and external events (quality cuts). Finally, events are classified into different samples.

\subsection{Quality cuts}
\label{sec:sel_qual}

\begin{figure}[h!]
	\centering
	\begin{minipage}[t]{0,49\columnwidth}
		Phase I \\
		\includegraphics[width=\columnwidth]{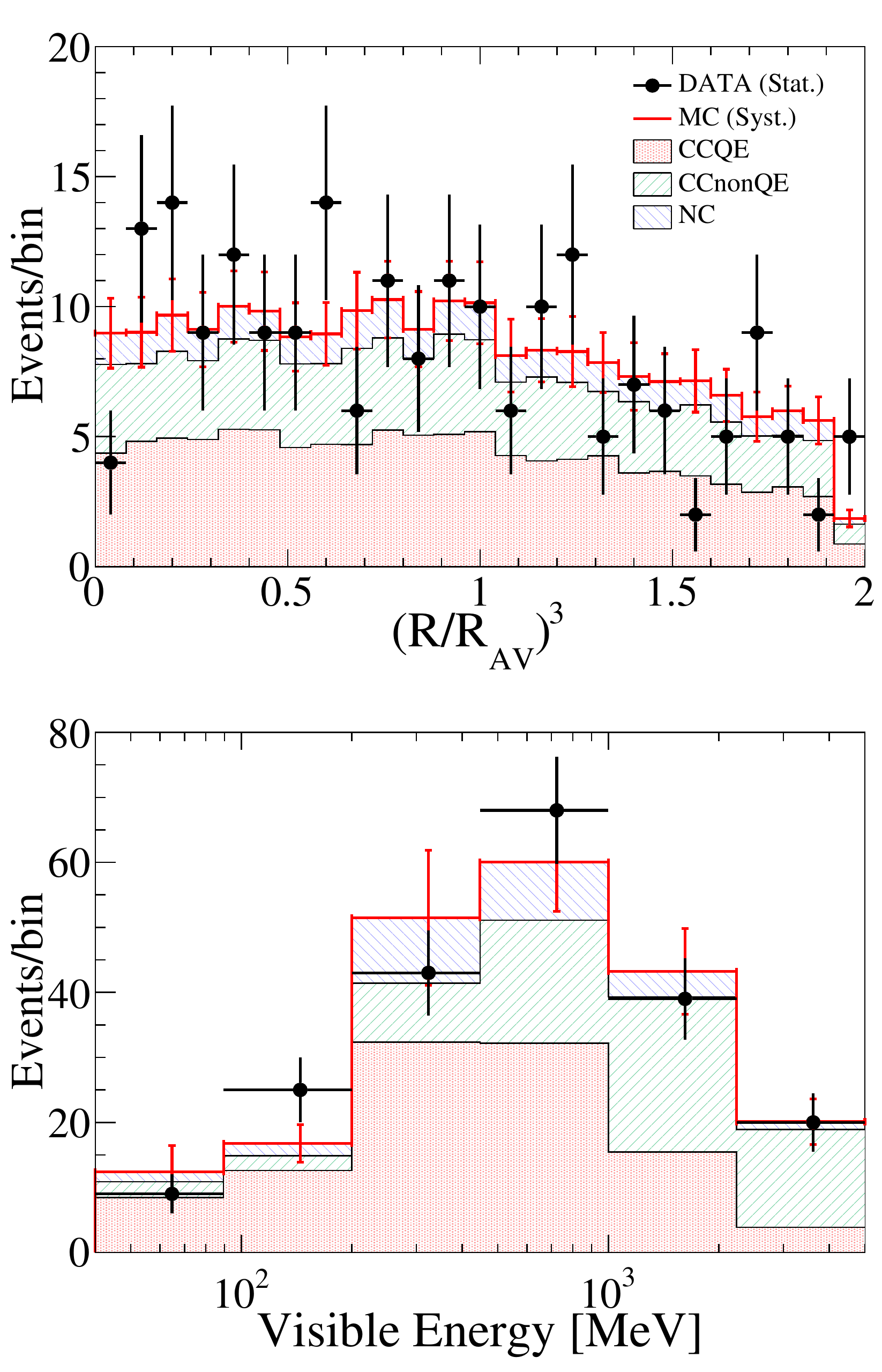}
	\end{minipage}
	\begin{minipage}[t]{0,49\columnwidth}
		Phase II \\
		\includegraphics[width=\columnwidth]{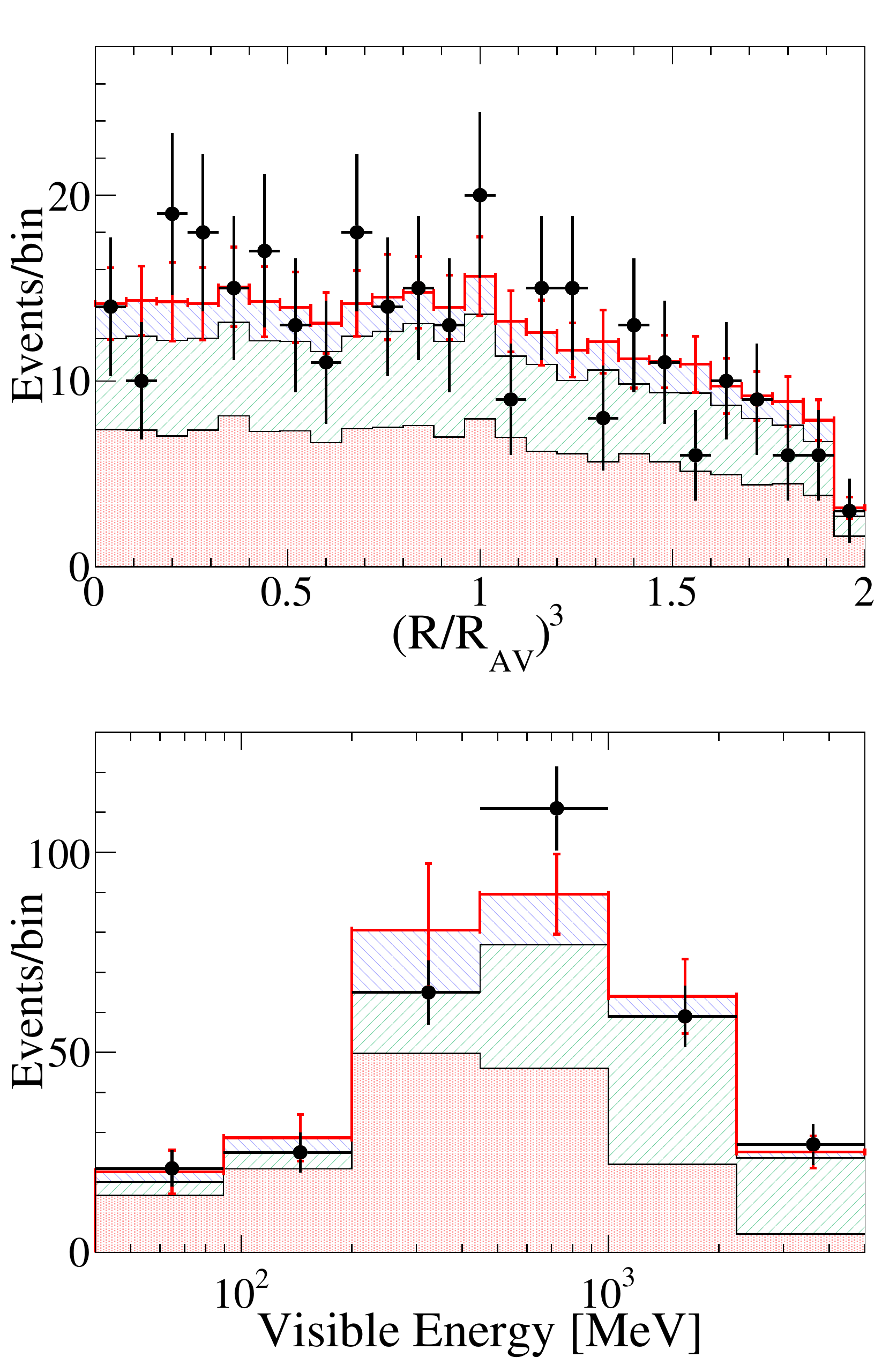}
	\end{minipage}
	\caption{\label{fig:sel} $\left( R/R_{AV} \right)^3$ (top) and visible energy (bottom) of the selected neutrino interaction candidates for Phase I (left) and Phase II (right). Black points correspond to data with only statistical uncertainties, and red bars correspond to MC with systematic uncertainties, broken down by neutrino interaction.}
\end{figure}

We have designed a criteria to identify fully contained events, i.e. events of which the charged particles deposited their entire energy in the active volume of the detector. Our main backgrounds are external cosmic muons and instrumental events, both generating high NHits events. The former is eliminated by requiring fewer than three triggered OWLs. Events due to external light leaking into the detector were identified and eliminated by requiring that none of the NECK PMTs is triggered. Events due to random flashes of light created by the PMTs, electronic pickup or sparks produced by PMTs are largely reduced to less than \SI{1}{\%} of the final selection using dedicated low-level cuts relying on event topology and PMT charge and timing information. A spherical fiducial volume of less than \SI{7.5}{\m} radius is chosen, and events reconstructing at a larger radius are removed in order to eliminate events that reconstruct poorly, partially contained events, and the external cosmic muon contamination. A low-energy threshold of \SI{50}{\MeV} is also applied. This criteria result in 204 selected neutrino interaction candidates in Phase I and 308 in Phase II. The $\left( R/R_{AV} \right)^3$ distribution is shown in \Cref{fig:sel}, where $R$ is the reconstructed radial position and $R_{AV}$ is the radius of the acrylic vessel. The visible energy distribution is shown in \Cref{fig:sel}. The MC is normalized to match the number of selected atmospheric neutrino events in data in order to directly compare the shapes. The absolute MC normalization is irrelevant for this analysis.

\subsection{Event classification}
\label{sec:sel_class}

We divided the entire dataset into CCQE or non-CCQE and separately into $\nu_{\mu}$ or $\nu_e$. CCQE interactions are typically characterized by having a single charged particle in the final state. This would lead to single-ring events, so we rely on determination of ring multiplicity in order to enhance CCQE interactions (CCQE selection) or enhance CCOther and NC candidates (non-CCQE selection). For the former, we require a single-ring event within a reduced fiducial volume of \SI{6.5}{\m}, while for the latter, we require just a multiring event. Hence, there are some events selected by the quality cuts that do not fall in any category. The PID capabilities of the reconstruction algorithm that separates showerlike events and tracklike events is sufficient to identify $\nu_e$ and $\nu_{\mu}$ interactions. The total number of selected events and the fraction of each component are shown in \Cref{tab:selmc_all} for each selection.

\begin{ruledtabular}
	\centering
	\begin{table*}
			\begin{tabular}{c|c|c|c|c|c}
			Mode & Quality cuts & CCQE selection & Non-CCQE selection & Electronlike & Muonlike \\
			\hline
			\hline
			No. events (data) & 512 & 123 & 208 & 283 & 229 \\
			\hline
			CCQE   & 51.1(0.5)\%   & 64.5(1.2)\%  & 28.7(0.6)\%   & 47.4(0.7)\%   & 55.6(0.8)\% \\
			CCRES   & 22.1(0.3)\%   & 18.0(0.5)\%  & 29.1(0.5)\%   & 20.6(0.4)\%   & 23.9(0.5)\% \\
			CCDIS    & 13.3(0.2)\%   & 9.3(0.4)\%    & 19.9(0.4)\%   & 14.0(0.3)\%   & 12.5(0.3)\% \\
			CC Other & 0.18(0.02)\%   & 0.15(0.04)\%  & 0.34(0.05)\%   & 0.15(0.03)\%   & 0.21(0.04)\% \\
			NCES     & 0.23(0.03)\%   & 0.20(0.05)\%  & 0.23(0.04)\%   & 0.20(0.03)\%   & 0.26(0.04)\% \\
			NC Other & 13.1(0.2)\%   & 7.8(0.04)\%    & 21.7(0.4)\%  & 17.7(0.4)\%   & 7.5(0.2)\% \\
			\hline
			$\nu_e$        & 48.9(0.5)\%   & 50.2(1.0)\% & 49.4(0.8)\% & 74.9(0.9)\%  & 17.5(0.4)\% \\
			$\nu_{\mu}$ & 47.6(0.5)\%   & 47.7(1.0)\% & 44.9(0.7)\%  & 20.5(0.4)\% & 80.5(1.0)\% \\
			$\nu_{\tau}$ & 3.5(0.1)\%     & 2.1(0.2)\%   & 5.7(0.2)\%    & 4.6(0.2)\%   & 2.1(0.1)\% \\
			\end{tabular}
		\caption{\label{tab:selmc_all} Number of events selected in data by the different criteria for both phases together (top row) and fraction of interaction types and neutrino flavor in each selection together as calculated using MC. The quality cuts criteria select an inclusive sample of neutrino interactions; the CCQE criteria enhance CCQE events; the non-CCQE criteria enhance CCOther and NC events; and the electron- and muonlike criteria enhance the corresponding lepton type. Given the different FV cuts for CCQE and non-CCQE selections, some events do not fall in either of those two categories. The uncertainties in parentheses correspond to the MC statistical uncertainties.}
	\end{table*}
\end{ruledtabular}

\section{\label{sec:nsel} Selection of neutron captures}

To identify neutron capture candidates, we require an event with energy larger than \SI{4}{\MeV} within a certain fiducial volume and in time coincidence with the neutrino interaction candidate, described in previous section. The main backgrounds are $^8$B solar neutrinos, the high-energy tail of radioactive backgrounds, and events due to instrumental noise. The former two categories are eliminated by the coincidence criteria, and the latter is greatly reduced by the low-level cuts originally designed for the SNO analyses, which leave an accidental coincidence rate lower than \SI{0.025}{\%}, as measured using randomly generated detector triggers. Production of unstable isotopes with lifetime and energy of the order of the neutron captures (like $^{12}$B) are expected to be more than 2 orders of magnitude smaller than that of neutrons \cite{sk_cosmo}.

\begin{figure}[!h]
	\centering
	\begin{minipage}[t]{0.49\columnwidth}
		\includegraphics[width=\columnwidth]{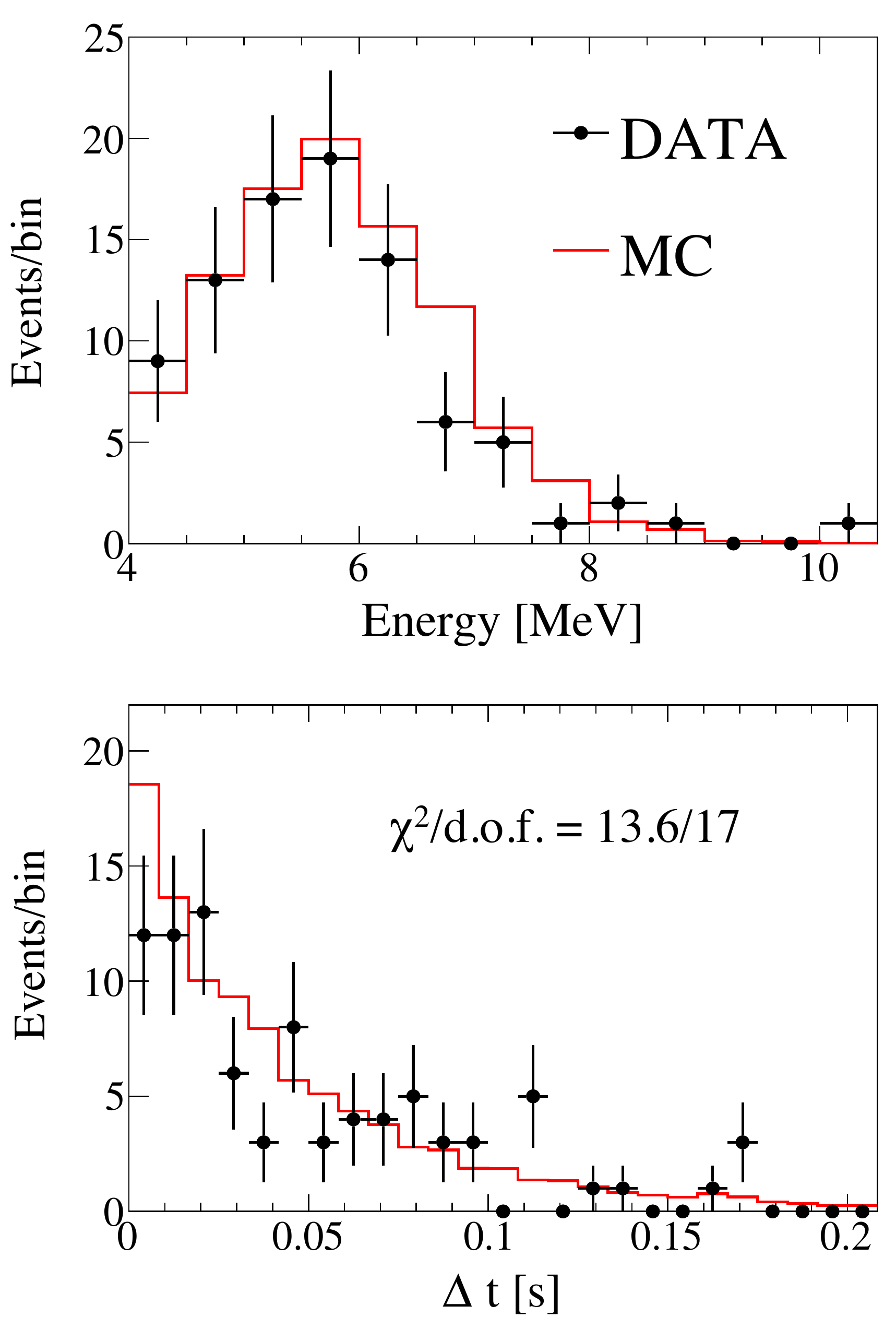}
	\end{minipage}
	\begin{minipage}[t]{0.49\columnwidth}
		\includegraphics[width=\columnwidth]{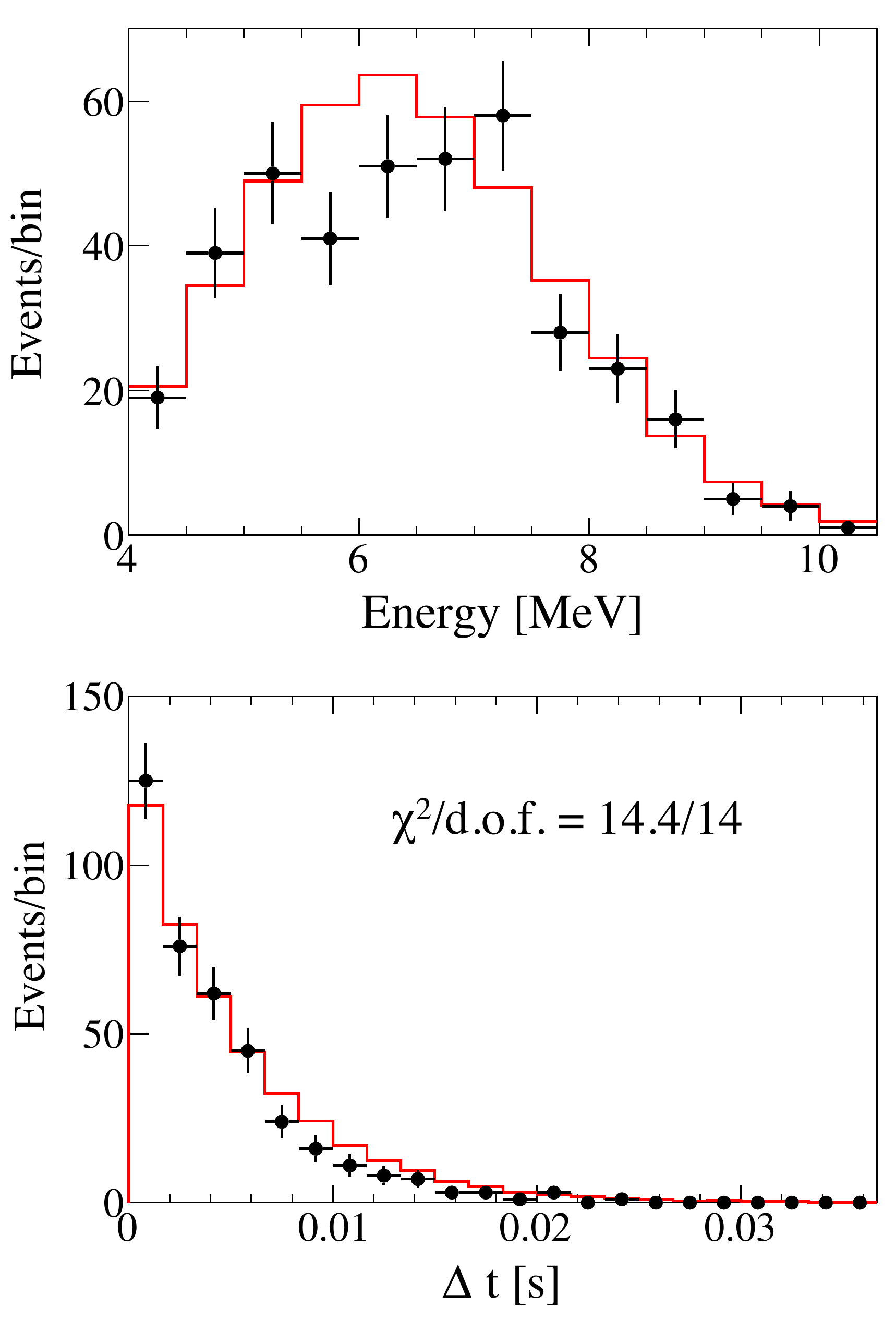}
	\end{minipage}
	\caption{\label{fig:sel_neut} Data and MC distributions for selected neutron captures candidates. Reconstructed neutron capture energy (top) and time difference with respect to neutrino interaction (bottom) for Phase I (left) and Phase II (right). MC is normalized by number of selected neutrino interactions. The $\chi^2$ values contain only statistical uncertainties.}
\end{figure}

We select all events within \SI{0.25}{\s} after an atmospheric neutrino candidate. Given that the neutron capture lifetime is of the order of a few milliseconds, the impact of this cut on the neutron detection efficiency is negligible. Events outside a fiducial volume defined by a sphere with \SI{6}{\m} radius are rejected. Random coincidences are largely mitigated by the \SI{4}{\MeV} energy cut. We confirmed through an independent analysis that the detector trigger efficiency is well modeled above \SI{4}{\MeV}. Finally, events with a $\Delta t < \SI{10}{\us}$ are rejected in order to eliminate possible Michel electrons and low NHit instrumental backgrounds. We select 88 neutron capture candidates in Phase I and 388 in Phase II. The energy distribution and the distribution of the time difference with respect to the neutrino interaction are shown in \Cref{fig:sel_neut} for both phases. The larger number of detected neutrons in Phase II is due the longer exposure and higher neutron detection efficiency with respect to Phase I. The MC is normalized to match the number of selected atmospheric neutrino events in data.

\begin{figure}[!h]
	\centering
	\includegraphics[width=\columnwidth]{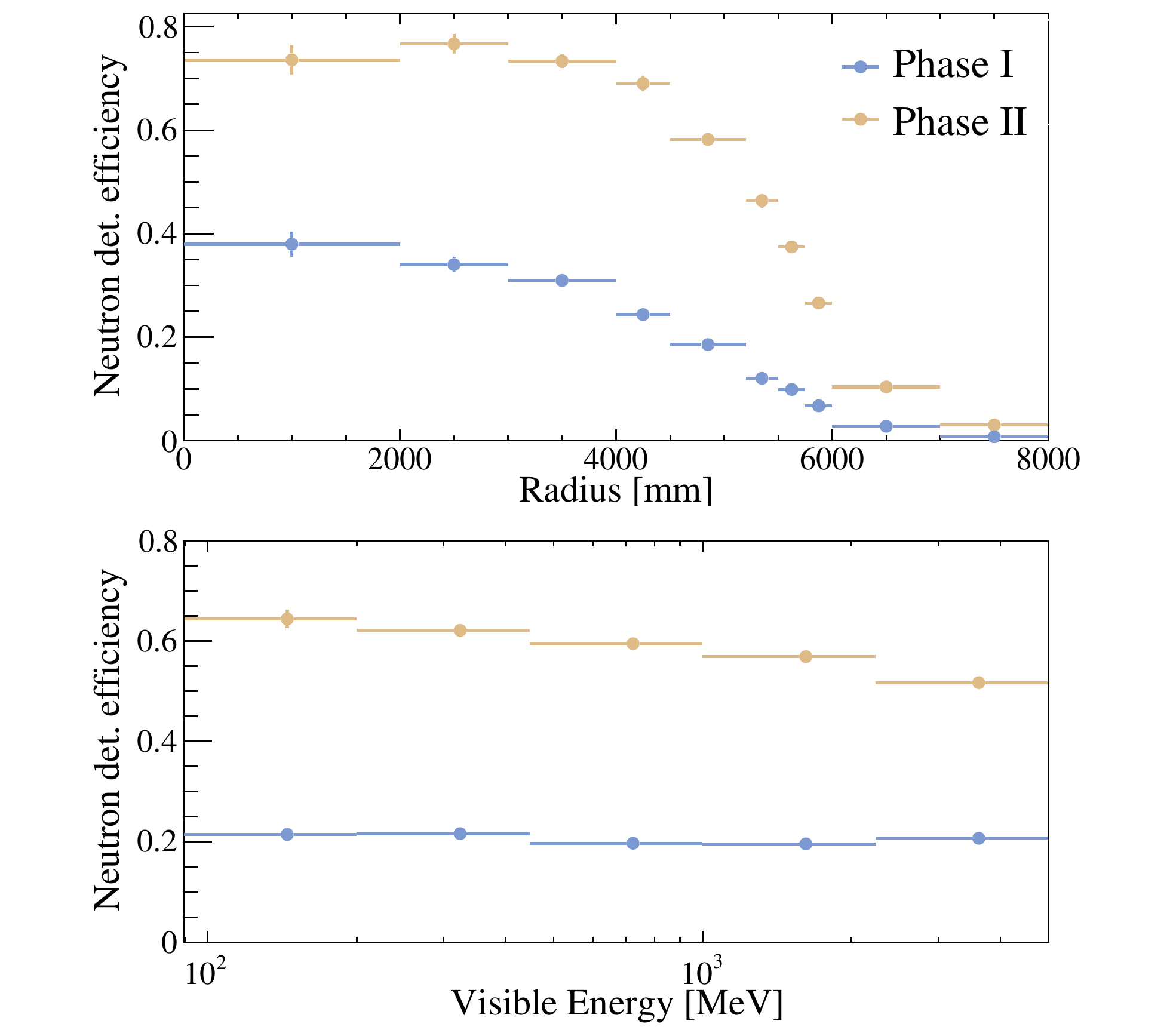}
	\caption{\label{fig:neff} Neutron detection efficiency as a function of the reconstructed neutrino interaction radial position (top) and the visible energy (bottom).}
\end{figure}

The total neutron detection efficiency was estimated from MC to be \SI{15.3}{\%} for Phase I and \SI{44.3}{\%} for Phase II. As shown in \Cref{fig:neff}, it features a strong dependency on the radial position of the neutrino interaction. This is due to the fact that neutrons created close to the light water (large radius) are more likely to leave the AV and capture in H, yielding a \SI{2.2}{\MeV} gamma ray, which is below detection threshold. The neutron detection efficiency increases significantly for Phase II, as expected. The plateau region near the center of the detector is due to the larger neutron absorption cross section of $^{35}$Cl as compared to $^2$H. The obtained efficiency values are compatible with the original neutron detection studies in \cite{sno_ndet}. The small differences are related to the fact that the energy of the neutrons produced by atmospheric neutrino interactions is typically higher than those produced by solar neutrinos, resulting in a higher chance of escaping the AV. The neutron detection efficiency decreases with energy since high energy neutrino interactions typically produce higher energy neutrons, which are more likely to exit the AV volume. In addition, the range of the particles produced in the neutrino interaction is larger at higher energies, so the production point of secondary neutrons could potentially be further from the neutrino interaction point inside the \DTO volume, and therefore be closer to the AV. The modelling of the neutron detection efficiency is studied using dedicated $^{252}$Cf calibration data (see \Cref{sec:ndet_eff}).

\section{\label{sec:syst} Estimation of systematic uncertainties}

A number of possible sources of systematic uncertainties are considered and estimated using various calibration sources and control samples. We separated them in the following categories: detector-related systematic uncertainties, cross section model uncertainties and uncertainties on the atmospheric neutrino fluxes and oscillation parameters. They are described in detail in the following sections.

\subsection{Detector systematic uncertainties}

\subsubsection{\label{sec:hecal} High-energy scale calibration}

In order to characterize the detector response at higher energies and calibrate the \textsc{Ring Fitter} energy reconstruction algorithm, data from two different sources were used: Michel electrons and stopping cosmic muons. The former provide an understanding of the intermediate energy scale since they provide a well-known energy distribution with a sharp cutoff at \SI{52.8}{\MeV}. The latter provide calibration of the \si{\GeV} energy scale since cosmic muons have a characteristic energy loss of approximately \SI{2.35}{\MeV\per\cm} in heavy water, so determination of the muon range provides a valuable calibration source for energies around approximately \SI{1}{\GeV}.

Michel electrons are easily identified by looking for events with more than 100 triggered PMTs preceded by an event in a time window between 0.7 and \SI{10}{\us}. Instrumental backgrounds are reduced by requiring that \SI{55}{\%} of the triggered PMTs are within a \SI{5}{\ns} window. PMT afterpulsing also occurs on timescales of a few microseconds and therefore could introduce an energy bias. The after-pulsing probability was determined to be \SI{1}{\%} per p.e. To mitigate after-pulsing contamination, only Michel electrons that are preceded by stopping muons with less than 2500 NHits are selected.

\begin{figure}[!h]
	\centering
	\includegraphics[width=0.98\columnwidth]{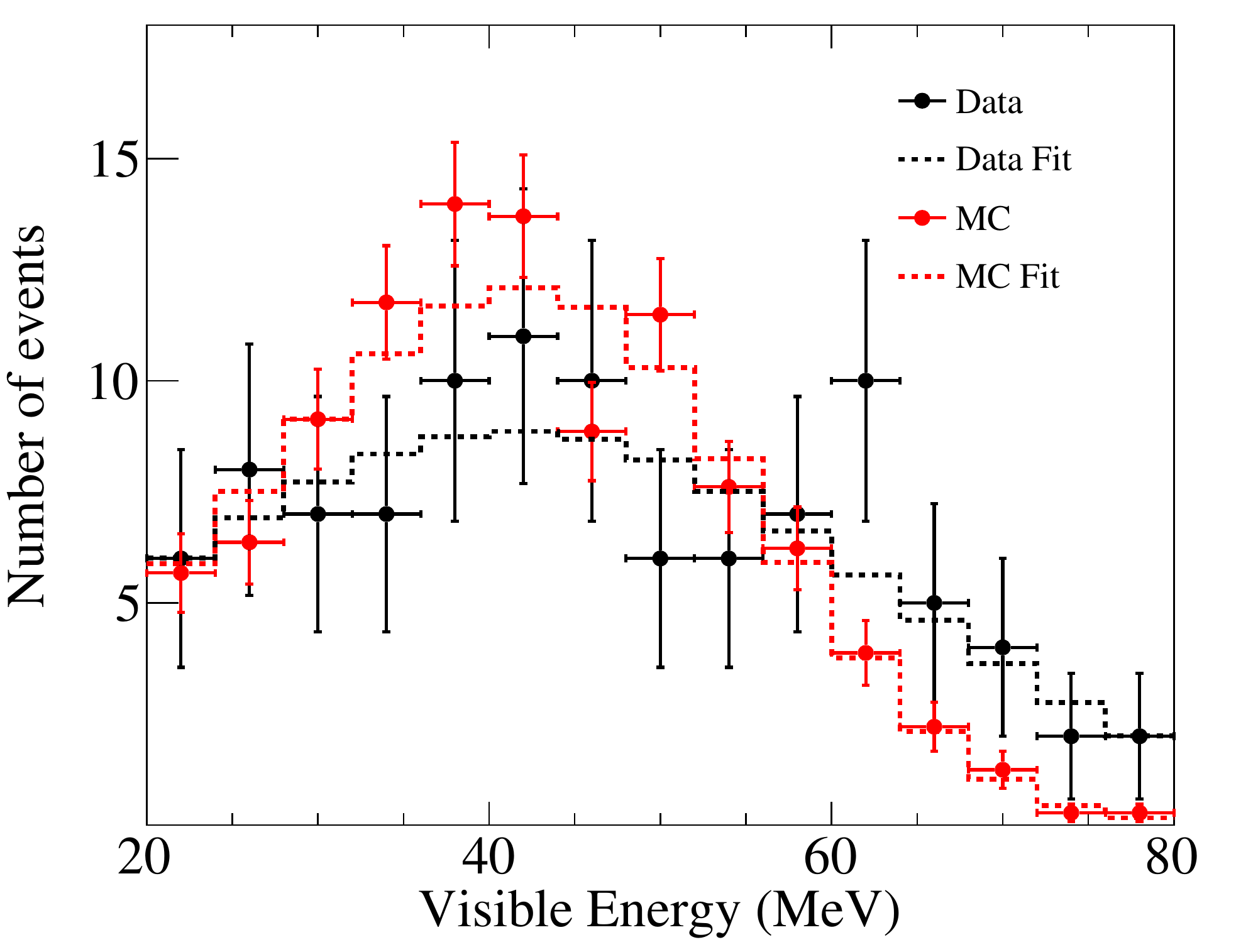}
	\caption{\label{fig:cal_michel_energy} Reconstructed energy distribution for the Michel electron control sample used for calibration and reconstruction benchmarking purposes. Points represent the data (black) and MC (red) reconstructed energy distributions. The dotted lines are the Michel electron fitted analytical expressions in \Cref{eq:michel}.}
\end{figure}

The Michel electron candidates are reconstructed using the \textsc{Ring Fitter} (\Cref{sec:rf}) and the visible energy distribution is fitted with the expected analytical form \cite{pdg18}
\begin{align}
\label{eq:michel} \left[3 \left( \dfrac{E + E_0}{E_M} \right) ^2 - 2 \left( \dfrac{E + E_0}{E_M} \right) ^3 \right] \otimes G(0,\sigma_E)
\end{align}
where $E$ is the energy which is constrained to $E<E_M$, $E_M=\SI{52.8}{\MeV}$ is the maximum permitted energy, $E_0$ is an energy shift correction and the last term represents a Gaussian smearing of width $\sigma_E$, which is interpreted as the energy resolution. The fit is done for data and for simulated Michel electrons generated using cosmic muons in \textsc{snoman} MC. The best fits are shown in \Cref{fig:cal_michel_energy} and correspond to an energy offset of \SI{4.1\pm4.1}{\MeV} for data and \SI{2.6\pm0.7}{\MeV} for MC with an energy resolution of \SI{18.9\pm4.7}{\MeV} for data and \SI{10.0\pm0.65}{\MeV} for MC. The energy bias is compatible between data and MC and the energy resolution for data is larger than predicted. The difference is attributed to the effect of unmodeled PMT after-pulsing, and to be conservative, it is propagated as a systematic uncertainty.

\begin{figure}[!h]
	\includegraphics[width=\columnwidth]{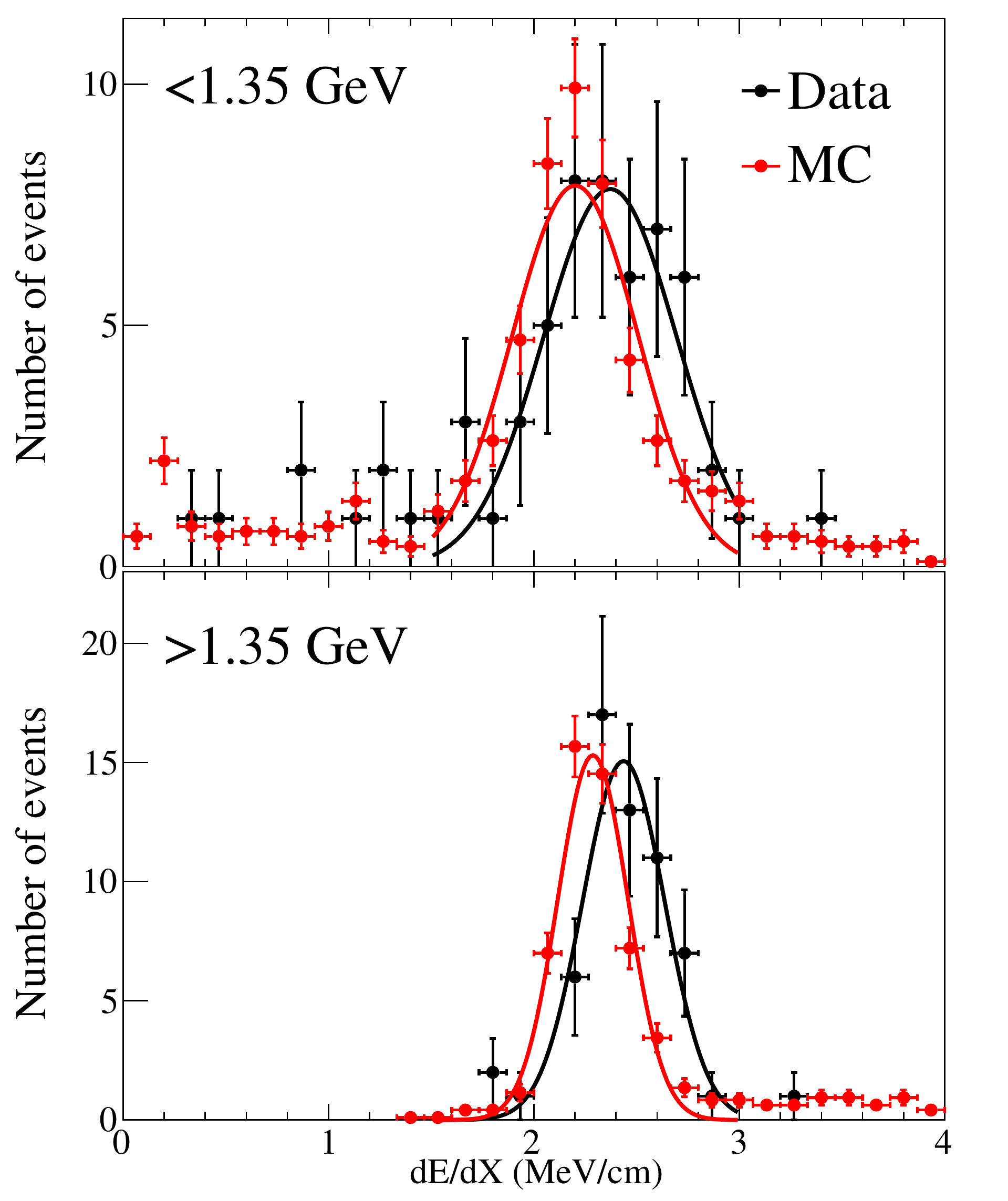}
	\caption{\label{fig:cal_muons_energy} Energy loss $dE/dX$ distribution for selected stopping cosmic muons for events with reconstructed energy below \SI{1.35}{\GeV} (top) and above \SI{1.35}{\GeV} (bottom).}
\end{figure}

External stopping muons produce a Michel electron signal near the end point of the track, allowing estimation of the muon range within the detector active region using the Michel electron's reconstructed position and the muon's reconstructed direction. Stopping cosmic muons are selected by requiring only one Michel electron candidate following an external event with more than three triggered OWLs. Since we are interested in single muon events, we reduce the dimuon and shower component by requiring a maximum of 25 triggered external veto PMTs. The neutrino-induced muon component is reduced by selecting downward-going events with $\cos\theta > -0.5$, where $\theta$ is the reconstructed zenith angle of the muon. We measure $dE/dX$ as the \textsc{Ring Fitter}-reconstructed muon-equivalent energy divided by the estimated muon range. The $dE/dX$ distributions are shown in \Cref{fig:cal_muons_energy}. We divided the dataset between low-energy (less than $1.35$~GeV) and high-energy (greater than $1.35$~GeV), in order to investigate any energy-dependent bias or resolution. Gaussian fits are performed for data and MC to estimate energy bias and resolution. The energy resolution is compatible between data and MC. We observe a small shift, which is attributed to a small difference in the averaged PMT gain used to reconstruct the energy along with possible misreconstruction of the muon track length. In the same fashion as was done with the Michel electron calibration, we err on the conservative side by propagating this difference as a systematic uncertainty.

\begin{figure}[!h]
	\centering
	\includegraphics[width=\columnwidth]{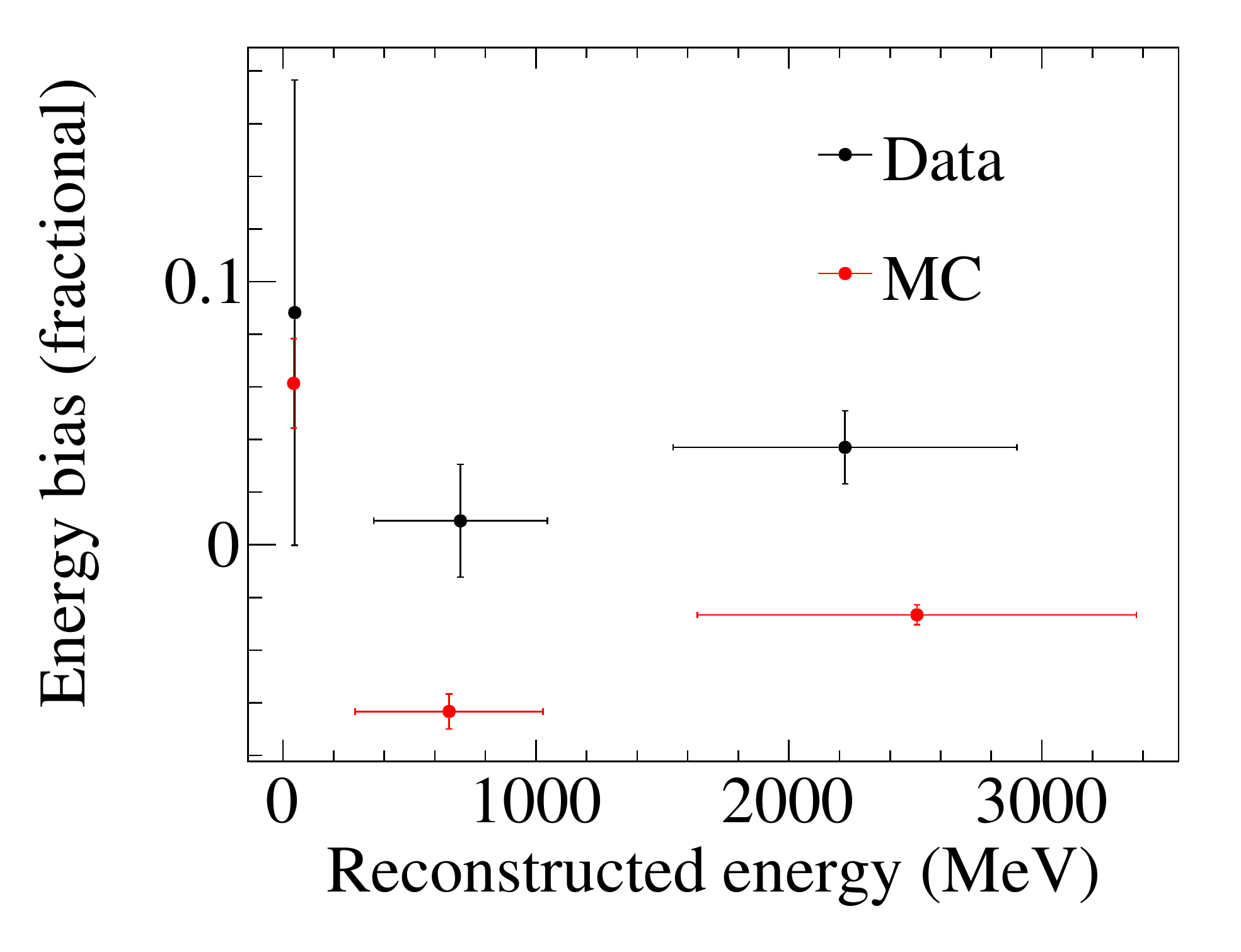}
	\includegraphics[width=\columnwidth]{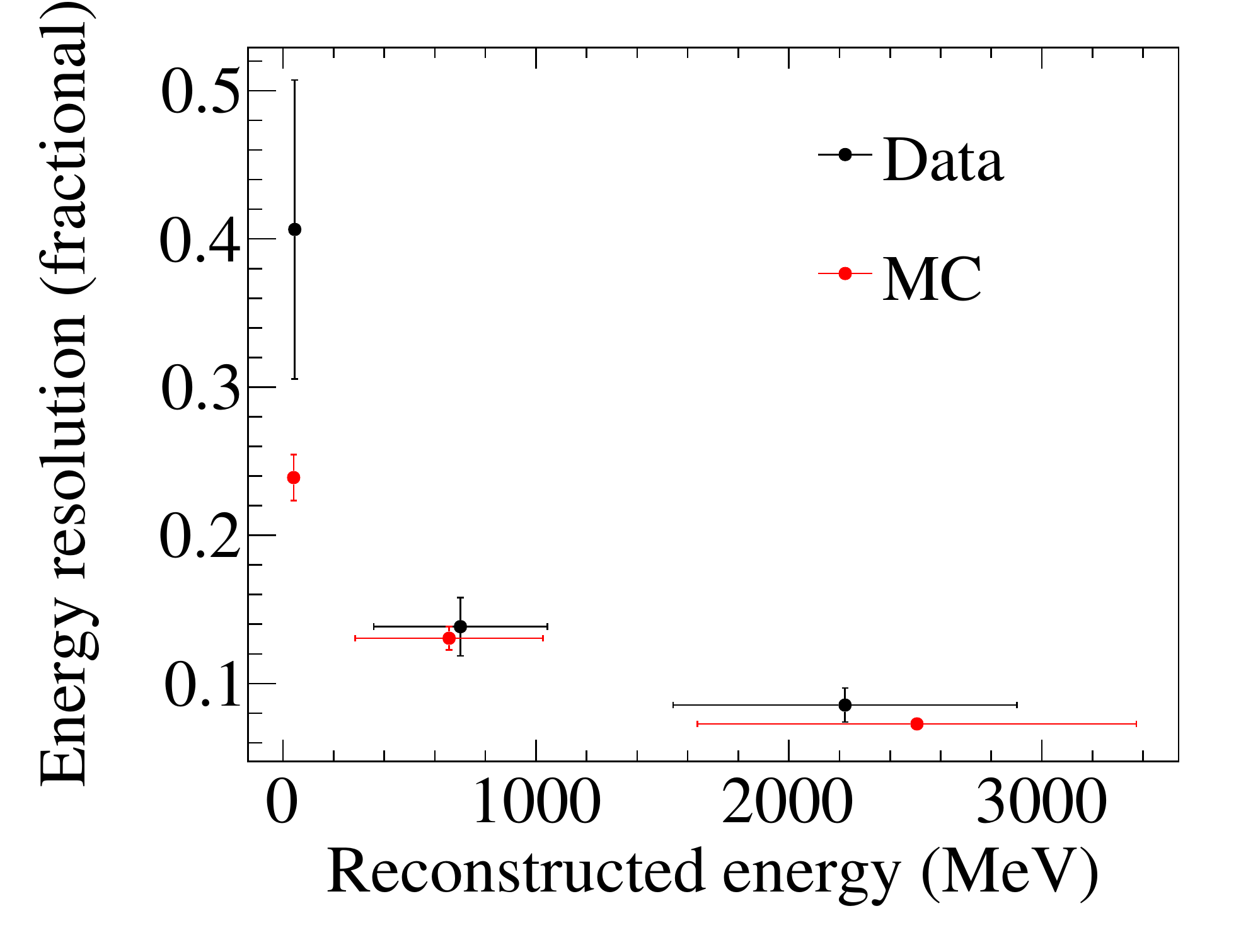}
	\caption{\label{fig:cal_energy_sum} Energy bias (top) and resolution (bottom) derived from the Michel electron (first point on the left) and stopping muon control samples (two last points on the right).}
\end{figure}

The summary of the final energy biases and energy resolutions is shown in \Cref{fig:cal_energy_sum}. The calculated energy bias is applied as a correction to data and MC. The differences between data and MC are propagated as a systematic uncertainty. To be conservative, the observed shift between data and MC is added to the uncertainty in the energy bias. The quadrature difference between the data and MC energy resolution is applied as a smearing to the MC, and the difference with the nominal MC is used to evaluate the systematic uncertainty.

\subsubsection{$E_{\nu}$ reconstruction}

The uncertainty in the angle between the incoming neutrino and outgoing lepton induces a systematic uncertainty in the reconstructed neutrino energy calculated from \Cref{eq:vreco}. The $1\sigma$ uncertainty is computed for every lepton energy bin, as shown in \Cref{fig:vreco_costh} and propagated into the final analysis.

\subsubsection{Atmospheric position bias and resolution}

External cosmic muons enter the detector through the spherical structure that holds the PMTs and, hence, at a specific known radius. This is used as a control sample to study performance of the radial position reconstruction for data and MC. Cosmic muons are selected as described in \Cref{sec:hecal}. An extra cut to remove events with more than 4000 triggered PMTs is applied, in order to have clearer rings and to ensure that no other effect could inflate the estimation of the systematic uncertainty. The \textsc{Ring Fitter} algorithm is applied to these events in order to reconstruct the entrance radial position $R$. The agreement of the reconstructed radial position between data and MC is good, with $\left\langle R^{dt}\right\rangle - \left\langle R^{mc}\right\rangle  = \SI{-28}{\mm}$, where $\left\langle R^{dt}\right\rangle$ and $\left\langle R^{mc}\right\rangle$ are the radial position averages for data and MC. The quadrature difference between the width of the radial distribution for data $\sigma_R^{dt}$ and MC $\sigma_R^{mc}$ is $\SI{160.0}{\mm}$.

\subsubsection{Particle identification and ring multiplicity performance}

We use the Michel electron and stopping muon candidate samples to test the performance on PID and ring multiplicity determination. The fraction of Michel electron events that misreconstruct as muon events is $11\pm1\%$ for MC and $7\pm3\%$ for data. For the stopping muons, $26\pm4\%$ are tagged as electrons for MC, in good agreement with $28\pm11\%$ for data. The difference is propagated in the analysis as a systematic uncertainty in the electron PID.

The rate of single particle events reconstructed as multiring events in the stopping-muon sample is $8\pm2\%$ for MC and $19\pm7\%$ for data. For the Michel electron sample, the number of events reconstructed as multiring corresponds to $1\pm0.2\%$ for MC and $15\pm7\%$ for data. These discrepancies are propagated into the analysis as a systematic uncertainty.

\subsubsection{Neutron capture energy and position systematic uncertainties}
\label{sec:syst_lene}

Reconstruction of the low-energy signal from neutron captures was extensively studied for the original SNO analyses \cite{sno_ndet}. The systematic uncertainties associated with the capture position, the position resolution, the energy scale and the energy resolution were computed using dedicated calibration campaigns where the different sources mentioned above were deployed. Comparison between MC and data yields the systematic uncertainties propagated in this analysis \cite{sno_leta}. The impact of these uncertainties is negligible compared to the rest of the systematic uncertainties.

\begin{ruledtabular}
	\begin{table*}
		\centering
		\begin{tabular}{c c c c}
			\textsc{genie} label & Physical parameter & Nominal value & 1$\sigma$ uncertainty \\
			\hline
			\hline
			\multicolumn{4}{c}{Cross sections} \\
			MaCCQE  & CCQE axial mass & \SI{0.990}{\GeV} & $-15\% + 25\%$ \\
			MaCCRES & CC and NC resonance axial mass & \SI{1.120}{\GeV} & $\pm 20\%$ \\
			MaCOHpi & CC and NC coherent pion production axial mass & \SI{1.000}{\GeV} & $\pm 50\%$ \\
			MvCCRES & CC and NC resonance vector mass & \SI{0.840}{\GeV} & $\pm 10\%$ \\
			R0COHpi & Nuclear size controlling pion absorption in Rein-Sehgal model & \SI{1.000}{\femto\m} & $\pm 10\%$ \\
			CCQEPauliSupViaKF & CCQE Pauli suppression via changes in Fermi level & \SI{0.225}{\GeV} & $\pm 35\%$ \\
			AhtBY, BhtBY & Higher-twist parameters in Bodek-Yang model scaling & $A=0.538, B=0.305$ & $\pm 25\%$ \\
			CV1uBY & GRV98 PDF correction param in Bodek-Yang model & $0.291$ & $\pm 30\%$ \\
			CV2uBY & GRV98 PDF correction param in Bodek-Yang model & $0.189$ & $\pm 30\%$ \\
			\hline
			\multicolumn{4}{c}{Hadronization} \\
			AGKYxF1pi & Pion transverse momentum in AGKY model \cite{agky}& \multicolumn{2}{c}{See Appendix C of Ref. \cite{genie}} \\
			AGKYpT1pi & Pion Feynman x for N$\pi$ states in AGKY model Ref. \cite{agky} & \multicolumn{2}{c}{See Appendix C of Ref. \cite{genie}} \\
			FormZone & Hadron formation zone & See Appendix C of Ref. \cite{genie} & $\pm 50\%$ \\
			\hline
			\multicolumn{4}{c}{Hadron transport} \\
			MFP\_pi, MFP\_N & Pion and nucleon mean free path & See Appendix C of Ref. \cite{genie} & $\pm 20\%$ \\
			FrCEx\_pi, FrCEx\_N & Pion and nucleon charge exchange probability & See Appendix C of Ref. \cite{genie} & $\pm 50\%$ \\
			FrAbs\_pi, FrAbs\_N & Pion and nucleon absorption probability & See Appendix C of Ref. \cite{genie} & $\pm 20\%$ \\
			\hline
		\end{tabular}
		\caption{\label{tab:syst_xsecs} Parameters adjusted in \textsc{genie} to estimate neutrino interaction systematic uncertainties. The parameters above the single horizontal line control the neutrino interaction cross section, while the ones below control the hadron transport models within the nucleus. See Ref. \cite{genie} for more details.}
	\end{table*}
\end{ruledtabular}

\begin{ruledtabular}
	\begin{table*}
		\centering
		\begin{tabular}{c c c c}
			Systematic parameter & $\pm1\sigma$ uncertainty & $1\sigma$ fractional effect & Type \\
			\hline
			\hline
			High-energy scale & \multirow{2}{*}{See \Cref{fig:cal_energy_sum}} & \multirow{2}{*}{$0.7\%$} & Shift \\
			High-energy resolution &  &  & Smearing \\
			Assumed $\cos \theta$ in $E_\nu$ reconstruction & See \Cref{fig:vreco_costh} & $<0.1\%$ & Shift \\
			\hline
			Particle misidentification & $e=0\pm5\%$, $\mu=4\pm5\%$ & $<0.1\%$ & Shift \\
			Ring miscounting & $e=14\pm14\%$, $\mu=11\pm9\%$ & $<0.1\%$ & Shift \\
			\hline
			High-energy radial bias & \SI{28}{\mm} & \multirow{2}{*}{$<0.1\%$} & Shift \\
			High-energy radial resolution & \SI{160}{\mm} & & Smearing \\
			\hline
			Quality cuts efficiency & 1.47\% & $1.5\%$ & Reweight \\
			\hline
			Neutron capture reconstruction & See \Cref{sec:syst_lene} & $<0.1\%$ & Shift, smearing \& reweight\\
			\hline
			Neutron detection efficiency & See \Cref{sec:ndet_eff} & 15.9\% & Reweight \\
			Atmospheric neutrino flux & $\sim15\%$ & 1.5\% & Reweight \\
			Neutrino interaction model & See \Cref{tab:syst_xsecs} & 12.5\% & Reweight \\
			MC statistical error & -- & 1.9\% & Reweight \\
			\hline
			Total & -- & 24.9\% & -- \\
		\end{tabular}
		\caption{\label{tab:syst_prop} Summary of the different systematic errors propagated into the analysis. The first column details the source of systematic uncertainty. The second column is the $1\sigma$ size of the propagated uncertainty or a reference to the relevant section if a single value cannot be given. The third column provides the $1\sigma$ variation on the total number of produced neutrons per neutrino interaction. The fourth column is the method used to propagate the systematic uncertainty (see the text for details).}
	\end{table*}
\end{ruledtabular}

\subsubsection{Neutron detection efficiency}
\label{sec:ndet_eff}

The neutron capture efficiency for low-energy neutrons is characterized by the calibrations performed with a $^{252}$Cf source for both phases. The source was deployed at different radial positions, and the detection efficiency was measured and compared to the original MC simulation. It was found to agree within $1.9\%$ for Phase I and $1.4\%$ for Phase II, demonstrating that the neutron modeling built into \textsc{snoman} is well understood. We compared our simulation in \textsc{geant4} to the one in \textsc{snoman} by comparing both models for single neutrons produced at different energies and reproducing the capture efficiency calculated for the $^{252}$Cf source. The estimated neutron detection efficiencies for both models agree within $1\%$ for energies below \SI{10}{\MeV} and within $3\%$ ($5\%$) for Phase I (II) at higher energies. To be conservative, we propagated the differences as systematic uncertainties by adding them in quadrature to the numbers extracted from the $^{252}$Cf calibration. The systematic uncertainty due to the detection efficiency for neutrons at energies relevant to this analysis is dominated by the width of the distribution at each energy and radius bin. The overall resulting systematic uncertainty is $15.9\%$ and is the dominant systematic.

\subsubsection{Quality cuts selection efficiency}

External cosmic muons are used as a control sample in order to estimate the efficiency loss of the cuts described in \Cref{sec:sel_qual}. Dark noise in the OWLs leads to valid events being rejected due to the OWL cut. This is estimated by measuring the OWL noise rate by randomly forcing the detector to trigger at a rate of \SI{5}{\Hz}. Only $0.27\%$ of the forced triggered events have more than one OWL hit, and the random coincidence of 3 OWLs is below $0.05\%$. We conclude that the loss in efficiency due to this effect is negligible. A similar study is applied to the NECK PMTs concluding that none of these effects has an appreciable impact. The inefficiency of the quality cuts for the cosmic muon sample is \SI{1.5}{\%} for data and \SI{2.1}{\%} for MC, being compatible within statistical uncertainties. The quadrature difference between these two values is propagated as a systematic uncertainty.

\subsection{Neutrino interaction model uncertainties}

The number of predicted primary neutrons depends on the interaction models. \textsc{genie} implements a system to vary the different parameters that impact neutrino cross sections and FSI. We change each relevant parameter by $\pm1\sigma$, returning a factor for every single event, which is applied as an individual event weight. In this way, we obtain the $\pm1\sigma$ boundaries for the number of predicted neutrons. The \textsc{genie} parameters of which the uncertainties have been propagated are shown in \Cref{tab:syst_xsecs}, classified in cross-section, hadronization or hadron transport model parameters. Their nominal values and $1\sigma$ uncertainty are also shown. For this work, we varied the axial and vector masses for the CCQE, CCRES, and NC interactions; the parameters in the Bodek-Yang model for DIS; the mean free path, absorption probability, and charge exchange probability for hadrons traveling through the nucleus; the parameters associated to the AGKY hadronization model \cite{agky}; and the one associated to the hadron formation zone. The uncertainty in the cross section model is the dominant of the three categories.

\subsection{Neutrino flux uncertainties}

Uncertainties on the neutrino production model and the neutrino oscillation parameters affect the theoretical prediction of the neutrino flux at SNOLAB. The model uncertainties are mostly driven by the uncertainty in the composition and energy spectrum of the primary cosmic-ray fluxes and the solar modulation. These are provided by the Bartol Collaboration \cite{bartol04}. Uncertainties relating to neutrino oscillation parameters are included using the uncertainties provided by the PDG18 \cite{pdg18}. In addition, the oscillations depend on the production point of the neutrino, the uncertainties of which are estimated in Ref. \cite{vprod_height} and included in the calculation of the oscillations.

The aforementioned parameters are shifted within $1\sigma$, generating a set of toy MC used to calculate the $1\sigma$ error bands of the neutrino energy spectra. Those boundaries are used to propagate the flux systematic uncertainties into the analysis by reweighting the different components and taking the difference with respect to nominal as the estimated effect of these uncertainties.

\subsection{Systematic uncertainties propagation and summary}

The overall strategy of propagating systematic uncertainties consists of defining parameters that control the different uncertainties and redoing the analysis for different values of these parameters. The difference with the nominal value is interpreted as the size of the effect of the specific systematic uncertainty. There are three types of parameters depending on the nature of the propagation:
\begin{enumerate}
	\item Shift: the parameter is shifted by $\pm1\sigma$.
	\item Smearing: the observable is smeared using a Gaussian of width equal to $1\sigma$.
	\item Reweight: the event is given a weighted value, which corresponds to a $\pm 1\sigma$ deviation from the nominal parameter.
\end{enumerate}
The considered systematic uncertainties are shown in \Cref{tab:syst_prop}, where the size of the $1\sigma$ uncertainty and its impact in the analysis are included, along with the propagation method. The fractional effect in \Cref{tab:syst_prop} corresponds to the $1\sigma$ variation on the total number of produced neutrons per neutrino interaction. Bin by bin uncertainties are considered in the final measurement.

\section{\label{sec:res} Results}

\begin{figure}[!h]
	\centering
	\includegraphics[width=\columnwidth]{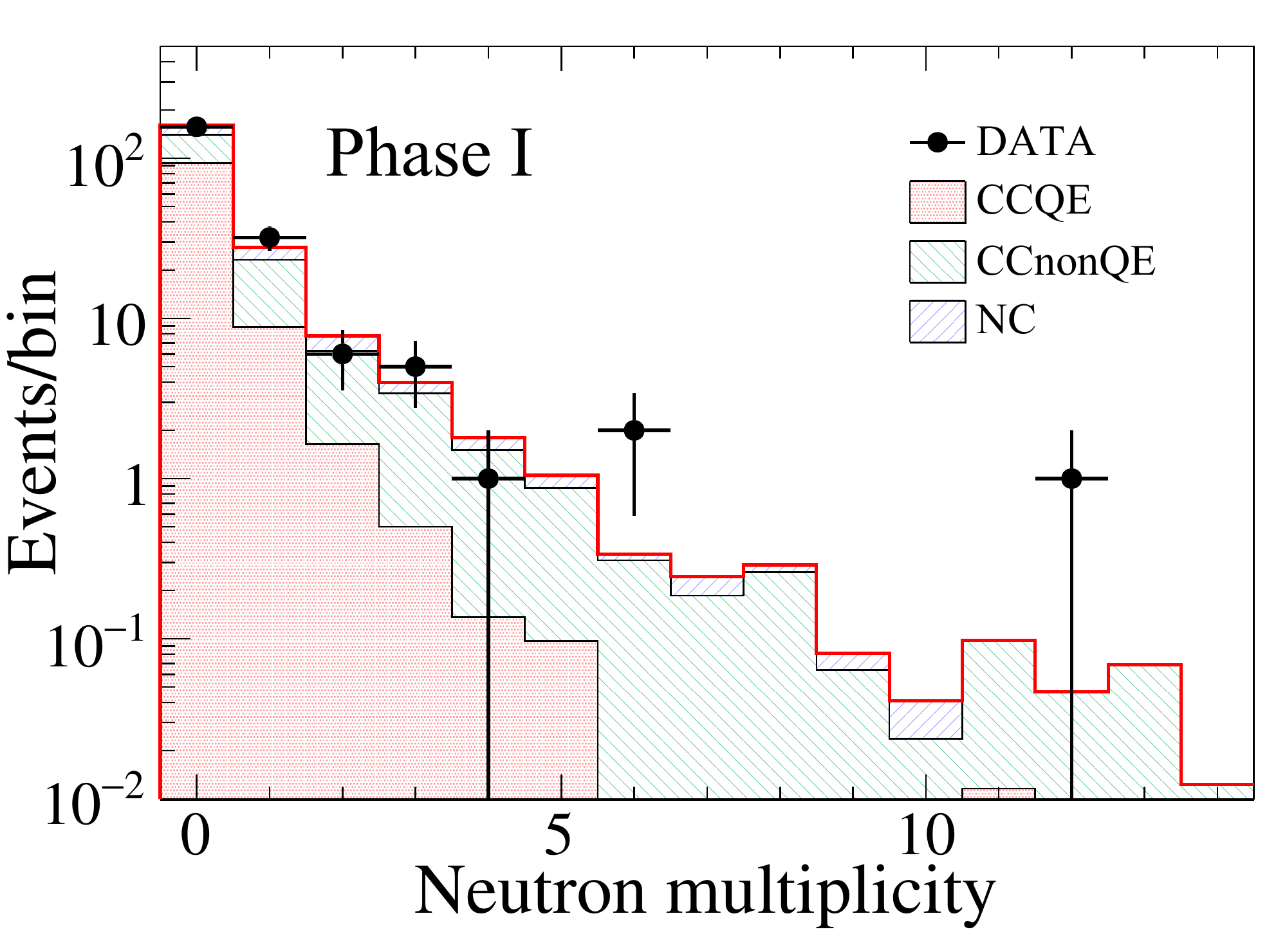}
	\includegraphics[width=\columnwidth]{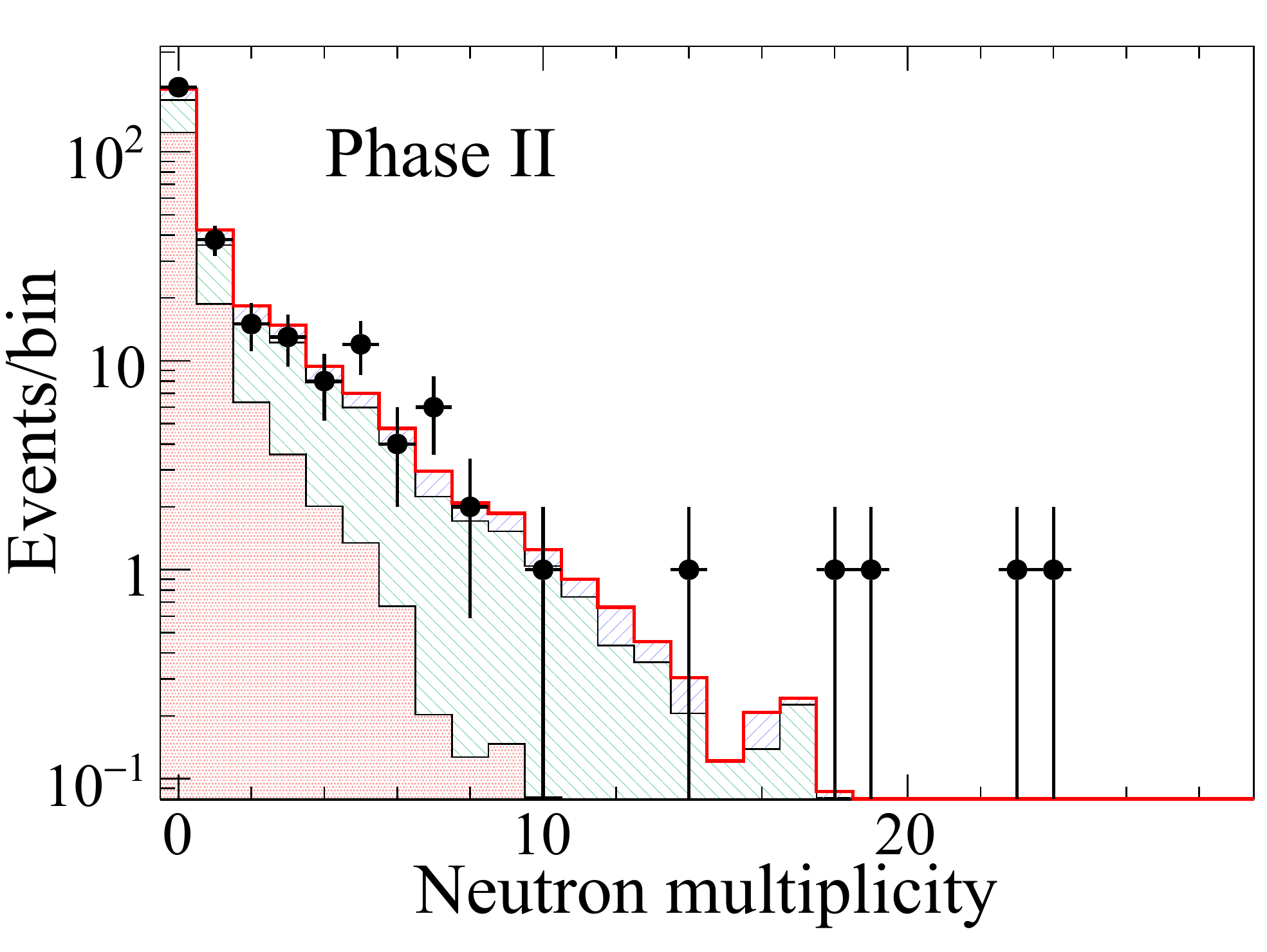}
	\caption{\label{fig:ndet} Number of detected neutrons per neutrino interaction candidate for Phase I (top) and Phase II (bottom).}
\end{figure}

The number of neutron capture candidates after an atmospheric neutrino interaction is shown in \Cref{fig:ndet} for both phases. The agreement between data and MC is good, although we identified four events with abnormally large neutron multiplicity in Phase II, compared to MC. Their energies and radial positions for the neutrino and neutron events are within the bulk of the population and the MC expectation.

\begin{figure}[h!]
	\centering
	\includegraphics[width=\columnwidth]{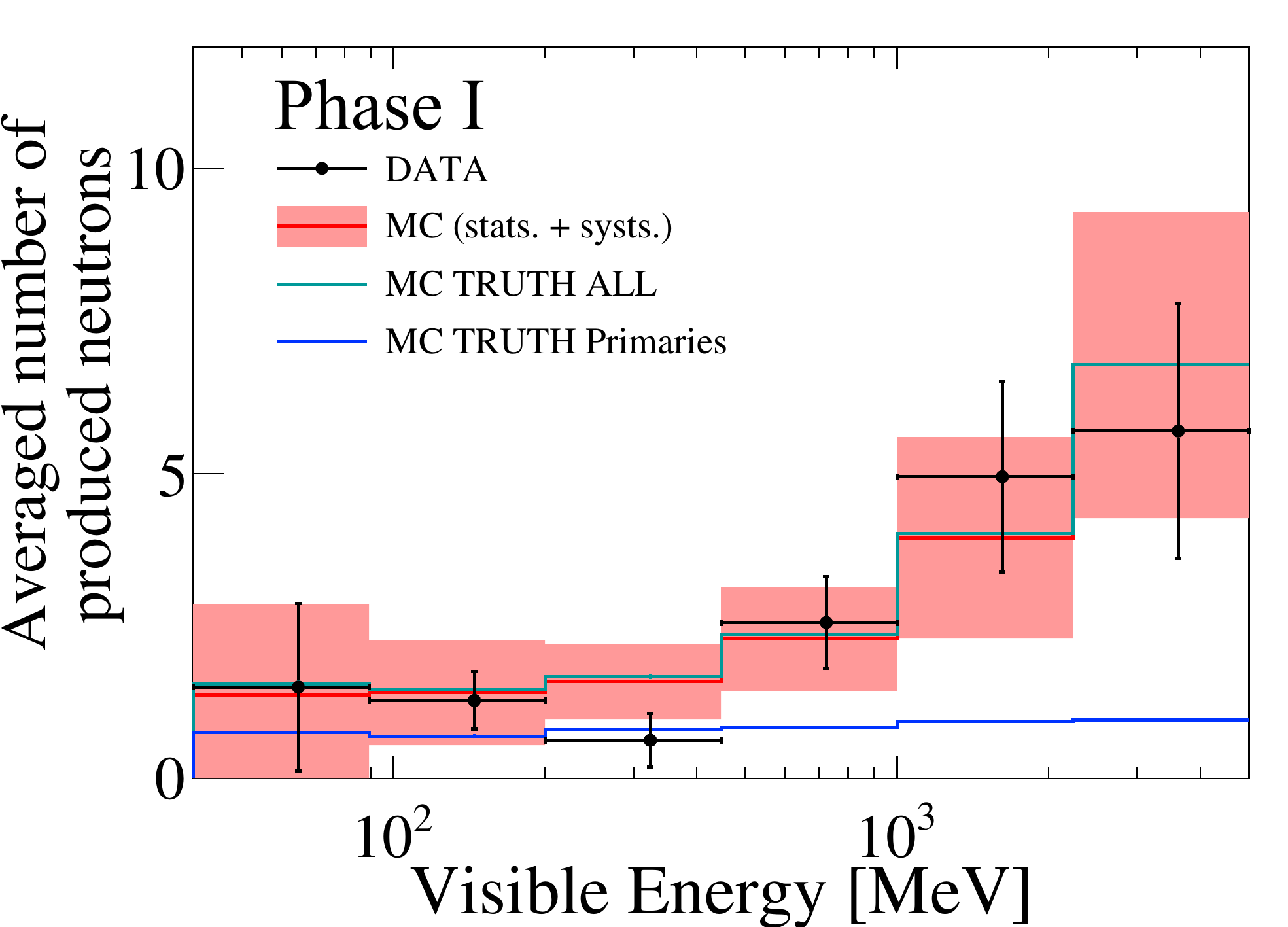}
	\includegraphics[width=\columnwidth]{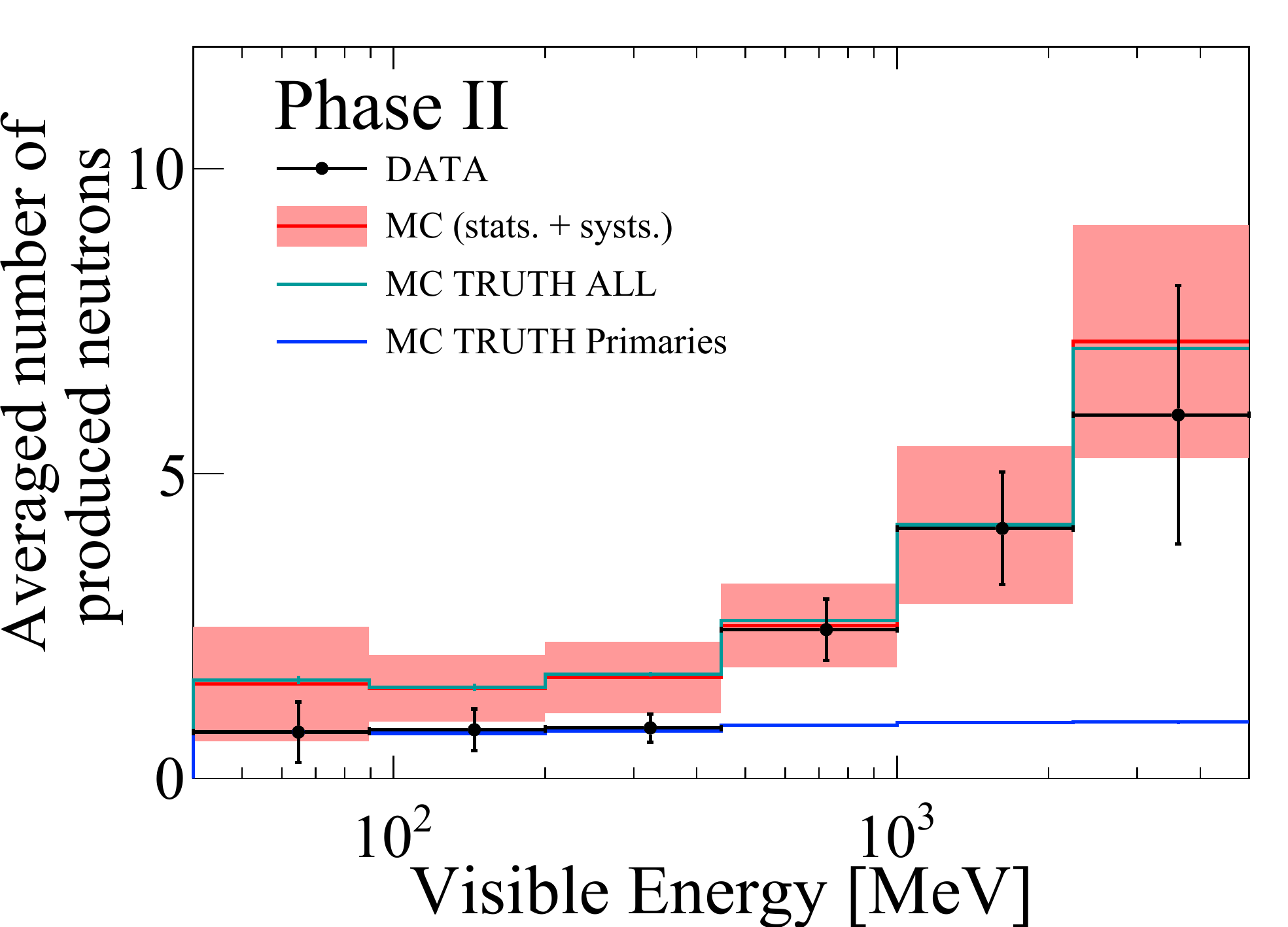}
	\caption{\label{fig:nprod_eEff} Averaged number of produced neutrons vs visible energy for Phase I (top) and Phase II (bottom). The points represent data with statistical uncertainties. The reconstructed MC is shown with red boxes with the size corresponding to the systematic uncertainties. The green line represents the average total number of neutrons given by the MC truth and the blue line corresponds to the average number of primary neutrons given by the MC truth.}
\end{figure}

After correcting for the calculated neutron detection efficiency shown in \Cref{fig:neff}, we estimate the average number of produced neutrons as a function of the visible energy in each phase, as shown in \Cref{fig:nprod_eEff}. The error bars on the data correspond to the statistical uncertainties while the size of the MC boxes represent the systematic uncertainties listed in \Cref{tab:syst_prop}. The $\chi^2/\mbox{ndof (number of degrees of freedom)}$ values are $8.17/6$ for Phase I and $10.8/6$ for Phase II, which include bin-to-bin correlations and correspond to p-values of $0.23$ and $0.09$, respectively. We performed a consistency check by comparing the efficiency-corrected neutron production in MC (red band) with the true neutron production (green line). This shows an excellent agreement, demonstrating that the efficiency correction is properly applied. The figure separates out the number of primary neutrons (blue line) to show how the production is dominated by secondary neutrons at higher energies, as discussed in \Cref{sec:nprod}. The measured neutron production shows good agreement between both phases, despite the different neutron detection efficiencies.

\begin{figure*}
	\centering
	\includegraphics[width=0.9\columnwidth]{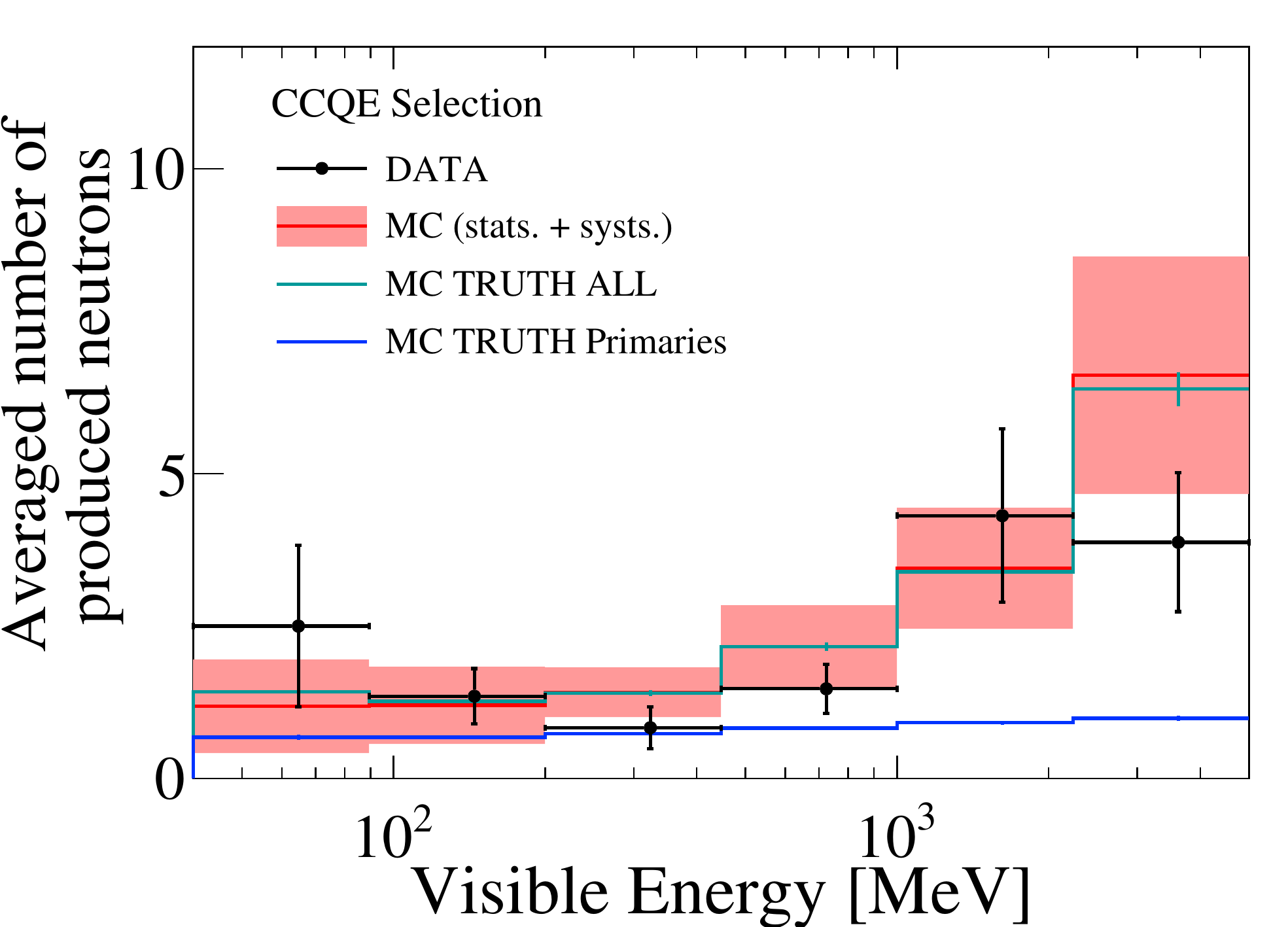}
	\includegraphics[width=0.9\columnwidth]{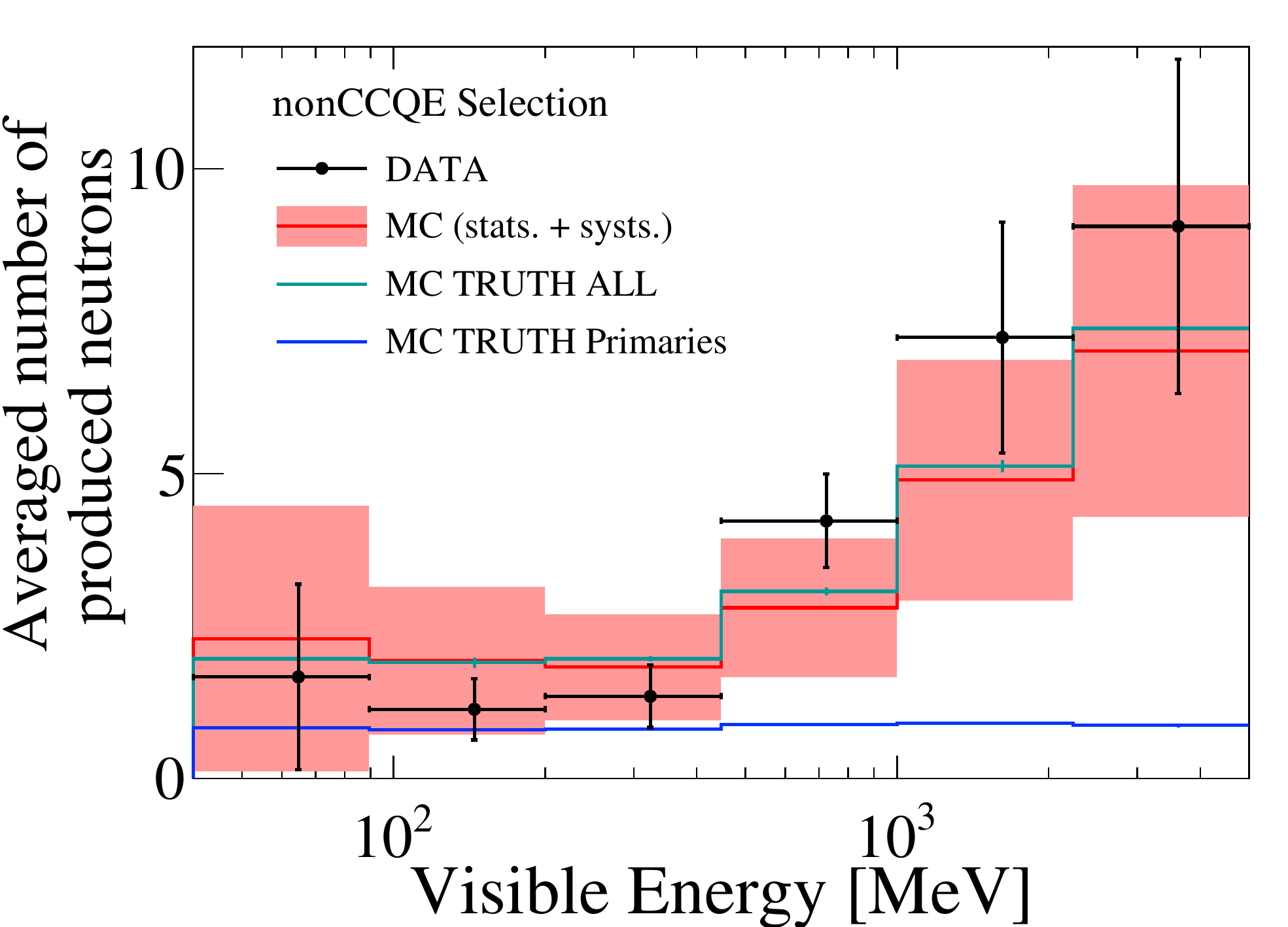}
	\includegraphics[width=0.9\columnwidth]{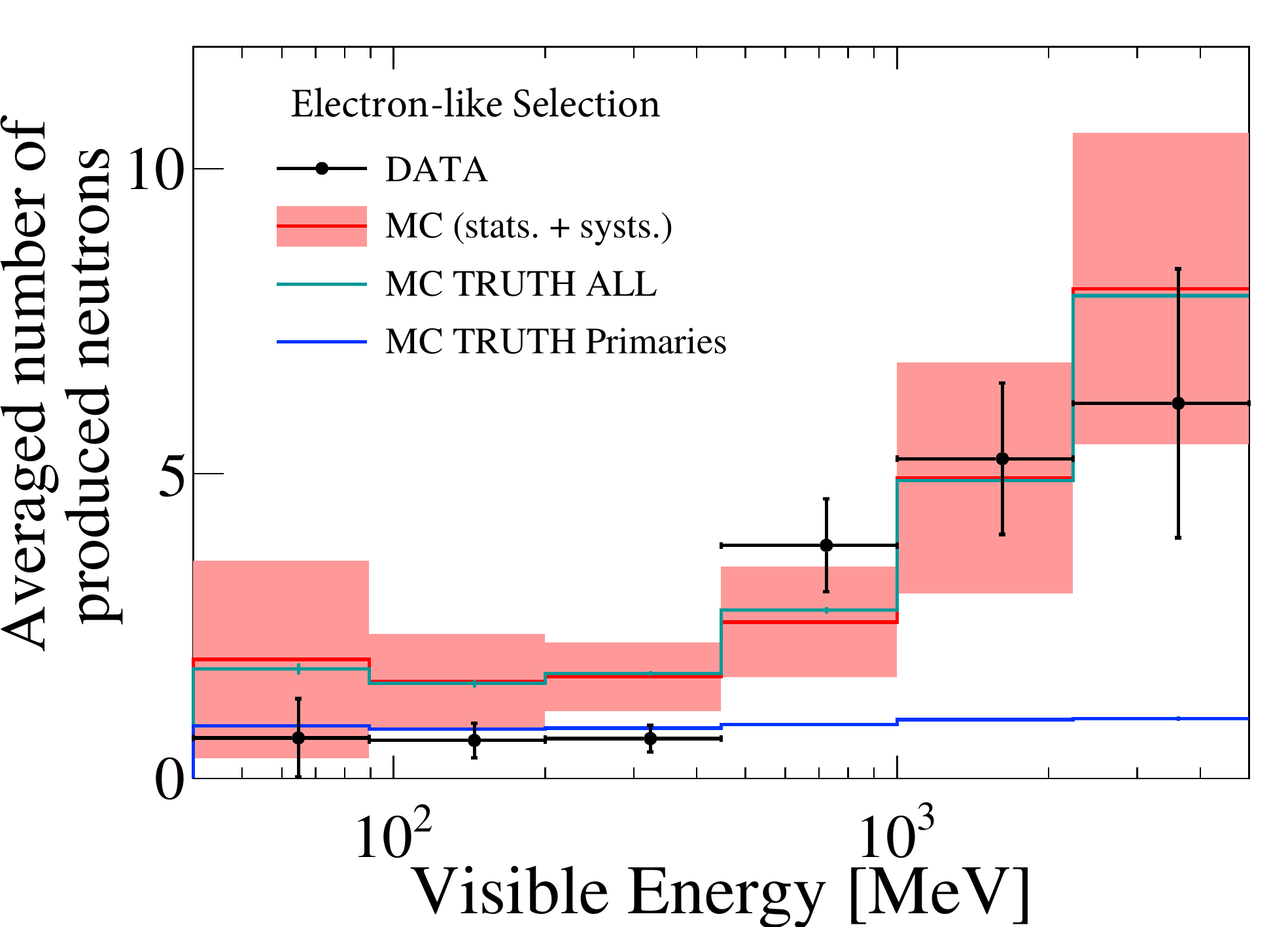}
	\includegraphics[width=0.9\columnwidth]{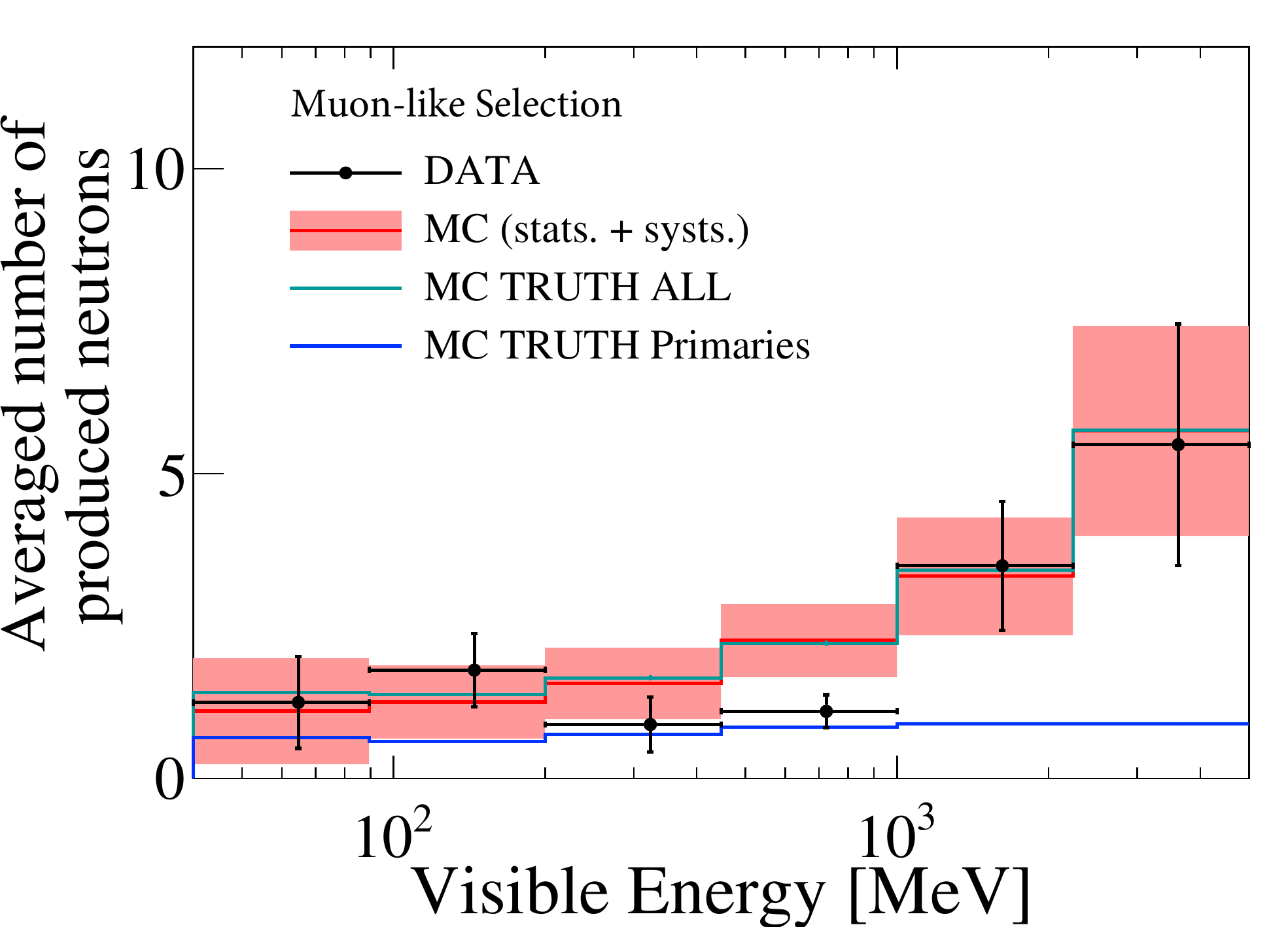}
	\caption{\label{fig:nprod_eEff_all_sels} Averaged number of produced neutrons vs visible energy for both phases together. We show the different selections: CCQE (top left), nonCCQE (top right), electronlike (bottom left) and muonlike (bottom right). The points represent data with statistical uncertainties. The reconstructed MC is shown with red boxes with the size corresponding to the systematic uncertainties. The green line represents the average total number of neutrons given by the MC truth, and the blue line corresponds to the average number of primary neutrons given by the MC truth.}
\end{figure*}

\begin{figure}[h!]
	\centering
	\includegraphics[width=0.5\textwidth]{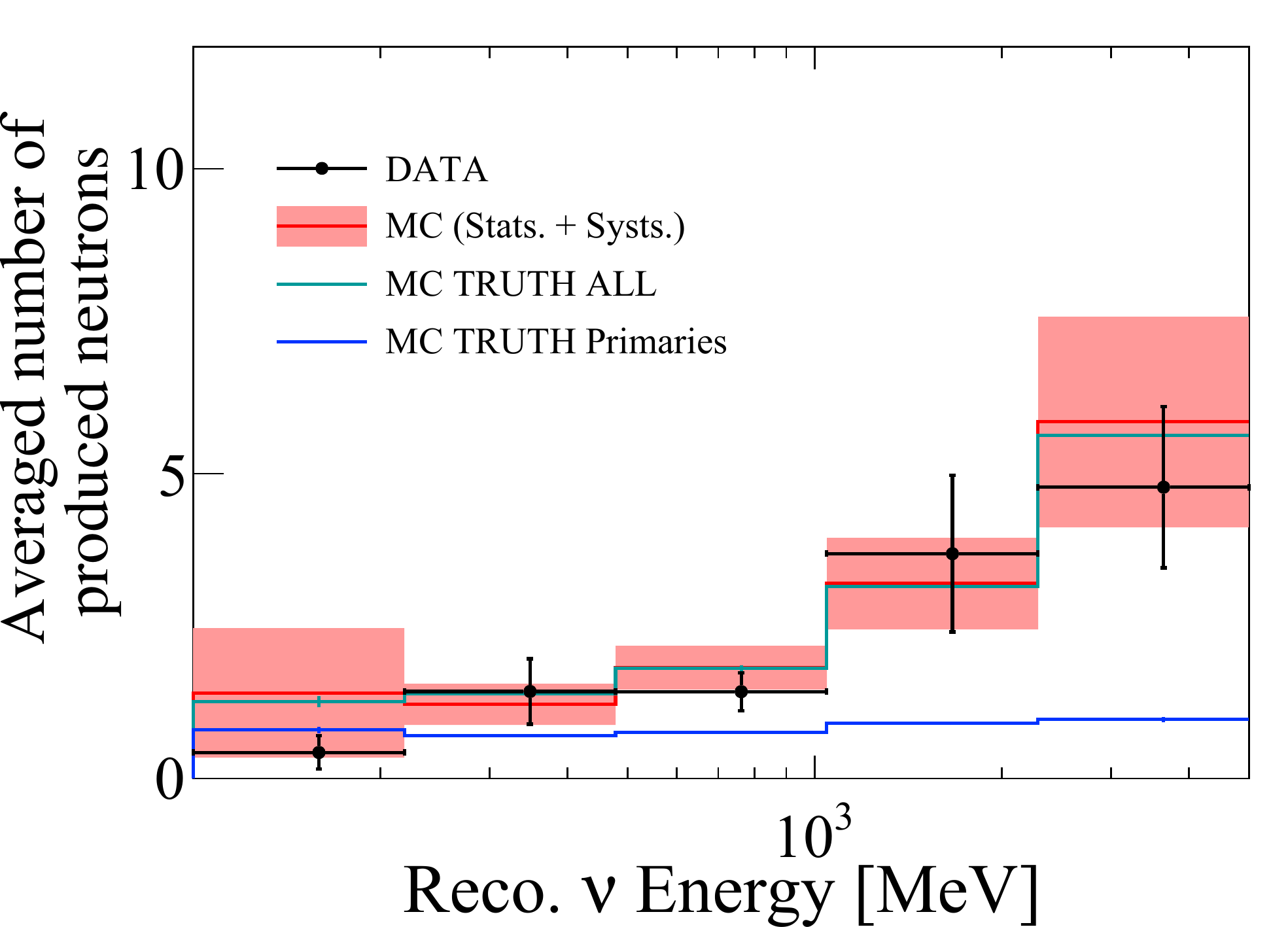}
	\caption{\label{fig:nprod_vE} Averaged number of produced neutrons versus reconstructed neutrino energy for both phases together for the CCQE selection. The points represent data with statistical uncertainties. The reconstructed MC is shown with red boxes with the size corresponding to the systematic uncertainties. The green line represents the average total number of neutrons given by the MC truth, and the blue line corresponds to the average number of primary neutrons given by the MC truth.}
\end{figure}

Based on the compatibility between phases, we performed an analysis on the combined dataset. The $\chi^2/\mbox{ndof}$ value on the average number of produced neutrons vs visible energy is $6.66/6$, which corresponds to a p-value of $0.35$. After classifying the full dataset as defined in \Cref{sec:sel_class}, the average number of produced neutrons is calculated and shown in \Cref{fig:nprod_eEff_all_sels} for each selection, allowing the study of neutron production for different interaction scenarios. The CCQE selection has a purity of $64.5\%$. For the non-CCQE selection, a purity of $71.3\%$ is achieved. Finally, the predicted neutron production for electronlike and muonlike events is overall in good agreement with the prediction. The neutrino energy is reconstructed for the CCQE-enhanced selection, and the neutron multiplicities are calculated with respect to this observable, as shown in \Cref{fig:nprod_vE}.

\begin{figure}[h!]
	\centering
	\includegraphics[width=0.5\textwidth]{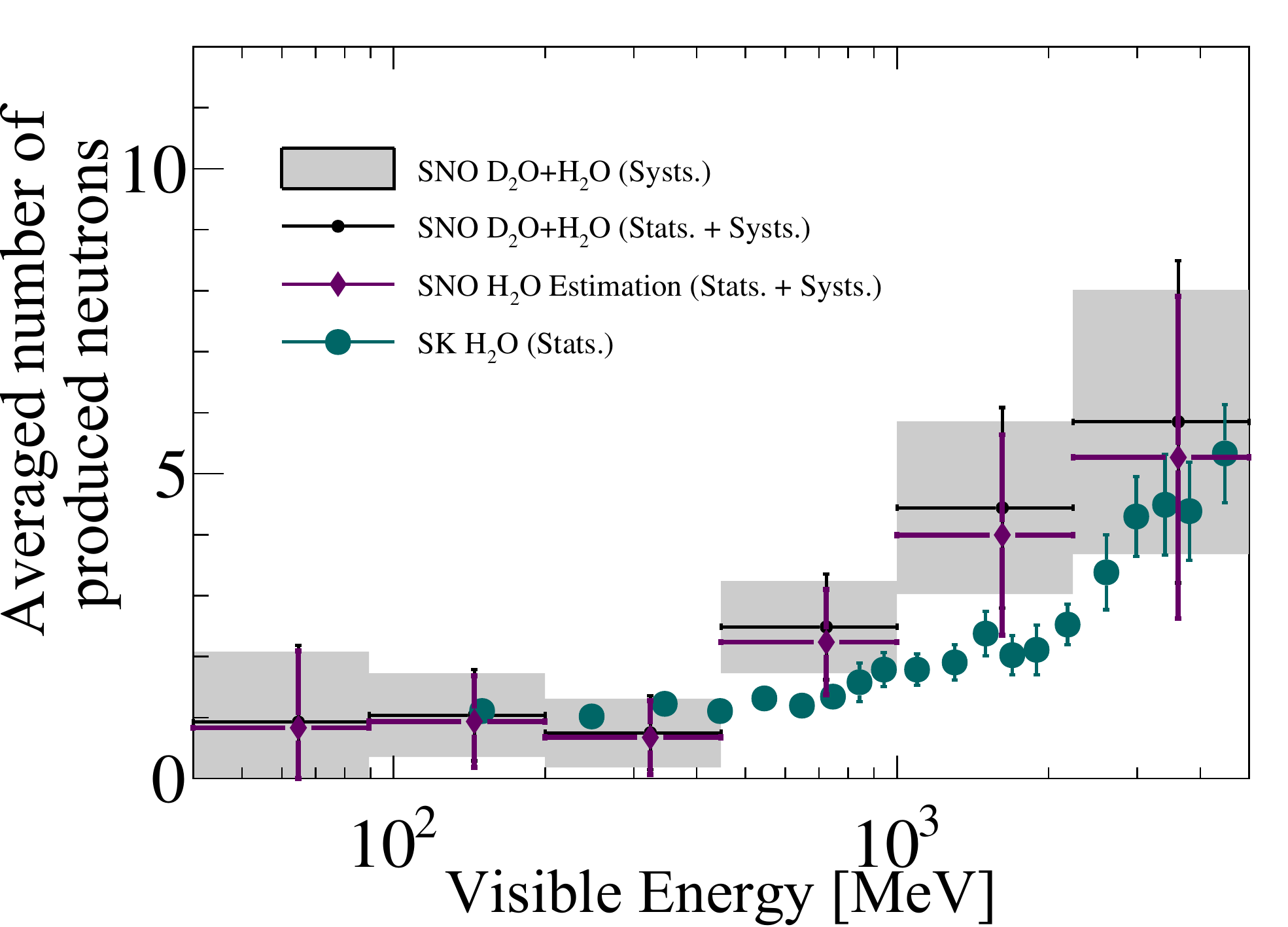}
	\caption{\label{fig:nprod_sk} Neutron production measurement in this work compared to SK published results \cite{superk_neutrons}. Black dots correspond to the present work, with gray boxes representing systematic uncertainties and solid lines being the total uncertainties. The estimation of SNO with pure light water (see the text for details) is shown with diamonds. The nominal SK measurement with light water is marked with circles, and it only displays statistical uncertainties.}
\end{figure}

We compared the total number of produced neutrons obtained by this work with the SK results \cite{superk_neutrons}. Since our measurement of neutron production is a combination of light and heavy water, we estimated the neutron production in a SNO detector filled with light water, in order to compare to the SK results. We calculate the expected neutron production difference between light water and heavy water by generating neutrino interactions in two SNO configurations: one with the AV filled with heavy water (nominal) and another with the AV filled with light water. \textsc{genie} vertices are produced in each geometry, and the final state particles are propagated in \textsc{geant4} as described in \Cref{sec:nprod}. According to our MC model, the total neutron production rate inside the analysis FV is $9.8\pm2.8\%$ larger for SNO with heavy water than for SNO with light water, driven by the larger production from neutron inelastic scattering. We estimated the neutron production in SNO with light water by scaling our measurement by $0.9$. In \Cref{fig:nprod_sk}, we show the comparison of the SNO measurement with the SNO with light water estimation and the nominal SK results. Our results are reasonably in agreement with SK data.

\subsection{Fit to primary and secondary neutrons}

\begin{figure*}
	\centering
	\includegraphics[width=\textwidth]{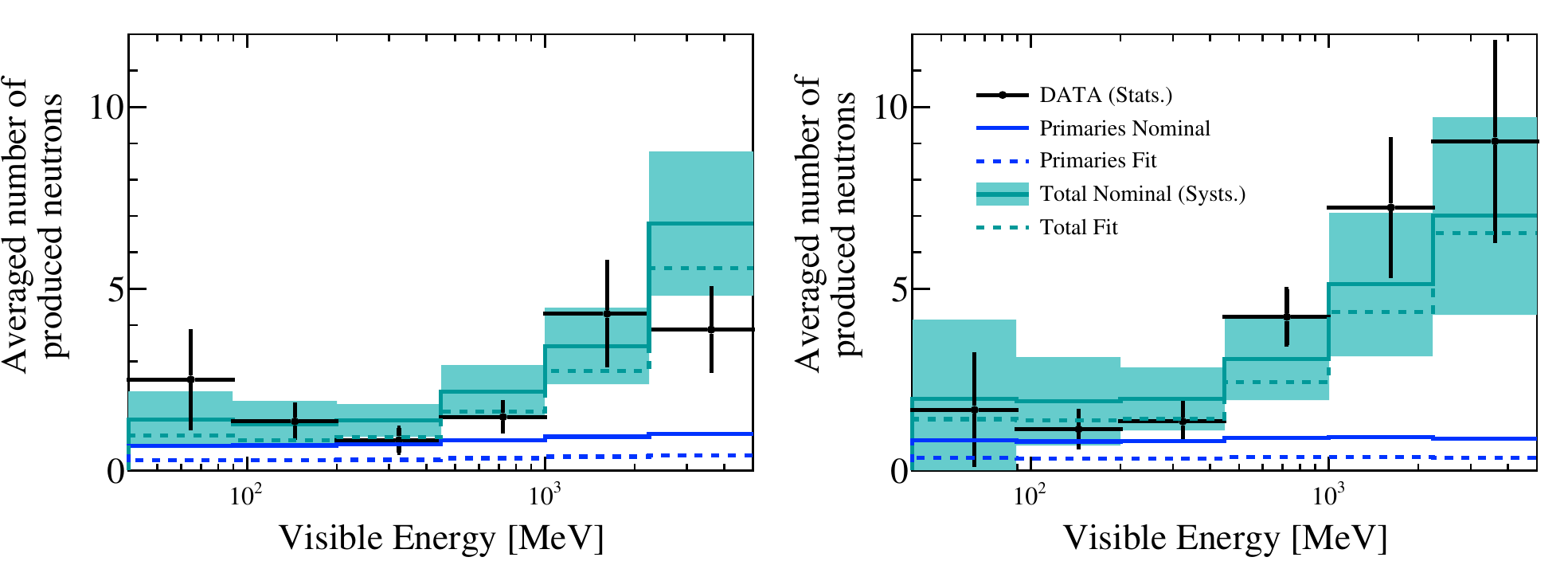}
	\caption{\label{fig:nfit} Number of produced neutrons vs visible energy for CCQE (left) and non-CCQE (right) selections before and after $\chi^2$ fit to neutron components. This combines both phases.}
\end{figure*}

The production of primary and secondary neutrons as a function of energy is very different---secondary neutrons production is larger at higher energy, while primary neutron production is rather flat (see \Cref{fig:nprod_venergy}). We estimate the contribution of each component by defining two normalization parameters (one for primary and another one for secondary neutrons) and constraining them with a $\chi^2$ fit. The difficulty of this analysis resides in the large correlations between these two parameters, given the uncertainties on the neutron production. We can break the degeneracy by fitting the CCQE and non-CCQE samples together, since the ratio between primary and secondary neutrons is quite distinct for CCQE and non-CCQE interactions (see \Cref{fig:nprod_int}). Before the fit, the nominal distributions show a p-value of $0.19$. The best fit for the normalization factors is $0.41\pm0.50$ for primary neutrons and $0.95\pm0.25$ for secondary neutrons, with a best fit $\chi^2/\mbox{ndof}=14.4/12$. The fit was performed using stand-alone CCQE, non-CCQE, electronlike and muonlike selections. The case presented here is the one that yields the lowest relative uncertainties. The uncertainty on the primary neutron production parameter is driven by a combination of the small production of primary neutrons and large uncertainties on the low-energy bins caused mainly by the neutron detection efficiency. \Cref{fig:nfit} shows the corresponding distributions before and after the fit. The difference with respect to the nominal prediction is small and features a p-value of $0.43$. The secondary production is compatible with the MC model prediction, while the fit prefers lower primary neutron production, being in slight tension with the nominal prediction. Similar fits to the different phases and selections yield compatible results. The systematic uncertainties described in \Cref{sec:syst} and the bin-to-bin correlations are taken into account in the fit.

\subsection{\label{sec:vvbar} Potential for $\nu-\bar{\nu}$ separation}

\begin{figure}[!h]
	\centering
	\includegraphics[width=\columnwidth]{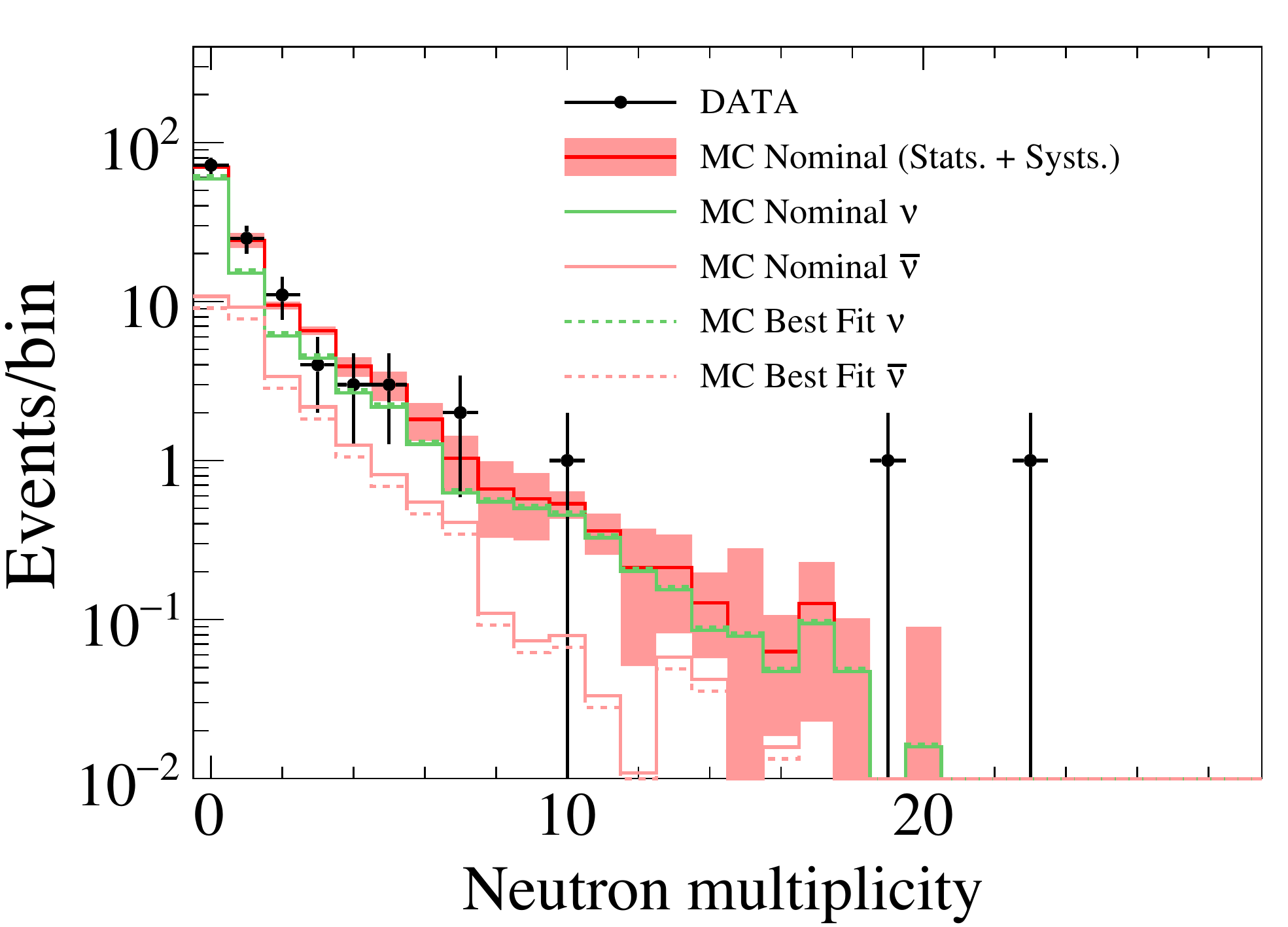}
	\caption{\label{fig:vvbarfit} Number of detected neutrons per neutrino interaction candidate in both phases. The MC shows the systematic uncertainties, and it is broken down into $\nu$ and $\bar{\nu}$. The fit results are also shown.}
\end{figure}

In \Cref{sec:nprod}, we showed how the simulation predicts that antineutrinos typically produce more neutrons than do neutrinos. This effect is enhanced in the CCQE case, since secondary neutron production is minimal, and antineutrinos produce on average one more primary neutron than do neutrinos (see \Cref{fig:nprod_int}). This feature is exploited to explore identification of neutrinos and antineutrino events by studying the distribution of the number of detected neutrons. Two normalization parameters are defined for the neutrino and antineutrino components and a $\chi^2$ fit is applied to the CCQE selection. The distributions before and after the fit are shown in \Cref{fig:vvbarfit}. It is important to notice the difference in shape between the two contributions, which breaks the degeneracy of the two components. We found a best fit value of $0.81\pm0.37$ for the normalization of the antineutrino component, in good agreement with the unity. This shows that we can constrain the antineutrino component at the $46\%$ level.

On the other hand, by selecting events with one or more detected neutrons, we enhance the number of antineutrino events from $23.6\%$ to $34.4\%$, according to the MC simulation.

\section{\label{sec:discussion} Summary and discussion}

We have measured the number of produced neutrons in atmospheric neutrino interactions as a function of the visible energy using the SNO detector. The neutrino interactions have been classified as $\nu_\mu$ vs $\nu_e$, and a subset has been classified as CCQE-like vs non-CCQE, in order to study the neutron production in each sample. The predictions from a MC model built using \textsc{genie} and \textsc{geant4} are in reasonable agreement with our measurements, although there are small tensions in certain energy regions. Data and MC are compatible within $2\sigma$ in the entire range and for every subsample. Comparison with published SK results\cite{superk_neutrons} shows a good agreement. We provided the neutron production as a function of the neutrino energy for CCQE events, showing that data and MC agree within $1\sigma$. We compared data to predictions of primary and secondary neutrons with a $\chi^2$ fit to the number of produced neutrons as a function of visible energy for the CCQE and non-CCQE selections. Our study of the separation of $\nu$ and $\bar{\nu}$ components using the number of detected neutrons shows that we can constrain the $\bar{\nu}$ component at the $46\%$ level and increase the purity of $\bar{\nu}$ events by $10.8\%$ by selecting neutrino events in coincidence with neutrons captures.

The projected future phase of SK with Gd-loaded water will be very interesting to better understand neutron production models. Furthermore, an experiment with larger statistics and higher neutron detection efficiency like ANNIE \cite{annie} will be very valuable to precisely study different neutrino-nucleus interactions and neutron production models as a function of interaction kinematics.

\clearpage

\begin{acknowledgments}
	This research was supported by Natural Sciences and Engineering Research Council of Canada; Industry Canada; National Research Council Canada; Northern Ontario Heritage Fund; Atomic Energy of Canada, Ltd.; Ontario Power Generation; High Performance Computing Virtual Laboratory; Canada Foundation for Innovation; Canada Research Chairs program (Canada); U.S. Department of Energy Office of Nuclear Physics; National Energy Research Scientific Computing Center; Alfred P. Sloan Foundation; National Science Foundation; the Queen’s Breakthrough Fund; Department of Energy National Nuclear Security Administration through the Nuclear Science and Security Consortium (United States); Science and Technology Facilities Council (formerly Particle Physics and Astronomy Research Council) (United Kingdom); Funda\c{c}\~{a}o para a Ci\^{e}ncia e a Tecnologia (Portugal). We thank the SNO technical staff for their strong contributions.  We thank INCO (now Vale, Ltd.) for hosting this project in their Creighton mine. We also thank the Super-Kamiokande Collaboration for allowing us to use their data in our comparison.
\end{acknowledgments}

\begin{thebibliography}{99}
	
		
\bibitem{t2k} The T2K Collaboration, Nucl. Instrum. Meth. A659, 106 (2011)

\bibitem{miniboone} The MiniBooNE Collaboration, Phys. Rev. Lett. 98 231801 (2007).

\bibitem{minerva} The Minerva Collaboration, Nucl. Inst. and Meth. A743 130 (2014).

\bibitem{nova} The NOvA Collaboration, NOvA Technical Design Report No. FERMILAB-DESIGN-2007-1.

\bibitem{dune} The DUNE Collaboration, Fermilab PAC, P-1062, (2015).

\bibitem{hyperk} The Hyper-Kamiokande Collaboration, Prog. Theor. Exp. Phys. 2015, 053C02 (2015)

\bibitem{sk_firstndet} The Super-Kamiokande Collaboration, Astroparticle Physics 31 320–328 (2009)

\bibitem{superk_neutrons} Super-Kamiokande Collaboration, Proceedings of the 32 International Cosmic Ray Conference, Beijing (2011)

\bibitem{genie} C. Andreopoulos et al, Nucl. Instrum. Meth. A614, 87 (2010).

\bibitem{genie_rw} C. Andreopoulos et al, arXiv:1510.05494[hep-ph].

\bibitem{geant4} S. Agostellini et al., Nucl. Instrum. Meth. A506 250-303 (2003)

\bibitem{sno_det} The SNO Collaboration, Nucl. Instrum. Meth. A449, 172 (2000).

\bibitem{snoman} Nucl. Instrum. Meth. A449 172 (2000)

\bibitem{bartol04} G. D. Barr et al., Phys. Rev. D 70, 023006 (2004).

\bibitem{n_geant4_fluka} Araujo \textit{et al.} \href{https://arxiv.org/pdf/hep-ex/0411026.pdf}{arXiv/0411026}

\bibitem{n_geant4_proton} M. Sabra, Nucl. Instrum. Meth. B358, 245 (2015).

\bibitem{ncap_cl36} B. Krusche \textit{et al.}, Nuclear Physics A 386, 245 (1982).

\bibitem{genie_hadronic} T. Yang \textit{et al.}, Eur. Phys. J. C 63 (2009).

\bibitem{richie_thesis} R. Bonventre, Publicly Accessible Penn Dissertations. 1213 (2014).

\bibitem{miniboone_fitter} R. B. Patterson \textit{et al.} Nucl. Instrum. Meth. A608, 206 (2009).

\bibitem{sk_fitter} A. D. Missert (The T2K Collaboration), J. Phys.: Conf. Ser. 888 012066 (2017).

\bibitem{hough} R. O. Duda and P. E. Hart, Commun. ACM 15, 11 (1972).

\bibitem{snop_nd} The SNO+ Collaboration, arXiv[hep-ex] 1812.05552 (2018)

\bibitem{root} Rene Brun and Fons Rademakers, Nucl. Inst. Meth. A389 (1997) 81.

\bibitem{E_b} E. J. Moniz \textit{et al.}, Phys. Rev. Lett., 26 445 (1971).

\bibitem{ftk} M. Dunford, PhD thesis. UMI-32-46154.

\bibitem{sno_n16} M.R. Dragowsky \textit{et al.}, Nucl. Inst. Meth A481 (2002) 284.

\bibitem{sk_cosmo} S. Li and J. Beacom, Phys. Rev. C 89, 045801 (2014).

\bibitem{sno_ndet} The SNO Collaboration, Phys. Rev. C 72, 055502 (2005) 

\bibitem{pdg18} The Particle Data Group, Chinese Phys. C Vol. 40, No. 10 (2016) 100001 

\bibitem{sno_leta} Phys. Rev. C 81, 055504 (2010)

\bibitem{agky} T. Yang \textit{et al.}, AIP Conf. Proc. 967 269-275 (2007)

\bibitem{vprod_height} T. W. Gaisser and T. Stanev, Phys. Rev. D 57(3) (1998)

\bibitem{annie} ANNIE Collaboration, arXiv:1402.6411v1 (2014)


\end{thebibliography}
\end{document}